# Atomic scale visualization of topological spin textures in the chiral magnet MnGe

Jacob Repicky[1], Po-Kuan Wu[1], Tao Liu[1,2], Joseph Corbett[1], Tiancong Zhu[1], Adam Ahmed[1], N. Takeuchi[3], J. Guerrero-Sanchez[3], Mohit Randeria[1], Roland Kawakami[1], and Jay A. Gupta[1]

[1]Department of Physics, The Ohio State University, Columbus, OH 43210, United States

[2] University of Electronic Science and Technology of China, Chengdu 610054, China

[3] Centro de Nanociencias y Nanotecnologia, Universidad Nacional Autónoma de México, Apartado Postal 14, Ensenada Baja California, Código Postal 22800, Mexico

**Abstract**

Spin polarized scanning tunneling microscopy is used to directly image topological magnetic textures in thin films of MnGe, and to correlate the magnetism with structure probed at the atomic-scale. Our images indicate helical stripe domains, each characterized by a single wavevector $Q$, and their associated helimagnetic domain walls, in contrast to the '3$Q$' magnetic state seen in the bulk. Combining our surface measurements with micromagnetic modeling, we deduce the three-dimensional orientation of the helical wavevectors and gain detailed understanding of the structure of individual domain walls and their intersections. We find that three helical domains meet in two distinct ways to produce either a 'target-like' or a 'π-like' topological spin texture, and correlate these with local strain on the surface. We further show that the target-like texture can be reversibly manipulated through either current/voltage pulsing or applied magnetic field, a promising step toward future applications.

Recent interest in topological spin textures in chiral magnets spans the range from fundamental science, such as Berry phase-induced Hall effects, to potential device applications including magnetic racetrack memories and neuromorphic computing (*1–8*). Spin textures such as helices and skyrmions result from competing magnetic interactions. In many cases of interest, the competition is between ferromagnetic exchange, favoring aligned spins, and the Dzyaloshinskii-Moriya interaction (DMI), which favors perpendicular spins, and which arises from spin-orbit coupling in the presence of broken inversion symmetry (*9*). The non-centrosymmetric 'B20' crystal structure of interest here breaks bulk inversion symmetry, and magnetic skyrmions were first discovered in B20 MnSi (*9*) and FeGe (*10*). In these materials, the magnetic phase diagram and its evolution from bulk crystals to thin films is now well understood. Within the B20 family, MnGe is an intriguing outlier (*11*), with a helical pitch length of 2.8 nm that is more than an order of magnitude smaller than other B20 crystals (*12*), and whose bulk phase diagram shows unusual "hedgehog-antihedgehog' crystals, for reasons that are not well understood (*13–15*).

We use spin-polarized scanning tunneling microscopy (SP-STM) to probe the magnetism on the surface of MnGe thin films with atomic resolution. SP-STM is uniquely suited to probe the rich physics of nanoscale spin textures in real space, and provides microscopic insights that complement those obtained from ensemble techniques that may average over different chiral domains (*16*). Our SP-STM images of 80 nm thick MnGe(111) films reveal a variety of topological spin textures depending on the local nanoscale structure. In atomically flat regions of the film, we find stripe-like contrast distinct from that expected for a 3$Q$ hedgehog lattice. The observed stripes with a 6 nm pitch can be understood as the surface projection of a 1$Q$ helical phase, where $Q$ is tilted with respect to the surface normal (*17*). In addition, we observed domain walls (DWs) between stripe domains and topological textures arising from intersecting DWs. We can understand these textures in detail using micromagnetic modeling that builds on recent advances in the theory of helimagnetic DWs (*18*). Specifically, we understand the rounding or sharpening of the helical stripes near DWs, and predict that there are only two characteristic ways in which three DWs can meet – the "target texture" and the "π-texture". Both of these textures are observed in the experiments and are correlated with local strain and curvature in the film. The target texture can be reversibly cycled through several distinct states by current/voltage pulsing and applied magnetic field.

Figures 1a-b show atomically-resolved topographic images of the MnGe(111) surface. These images are consistent with the B20 structure of MnGe, which features alternating Mn and Ge layers with atoms arranged in triangular lattices of single atoms or trimers (c.f., Supporting Information). The unit cell comprises three 'quadruple layer' (QL) repeats, and the relative stacking order and orientation of these

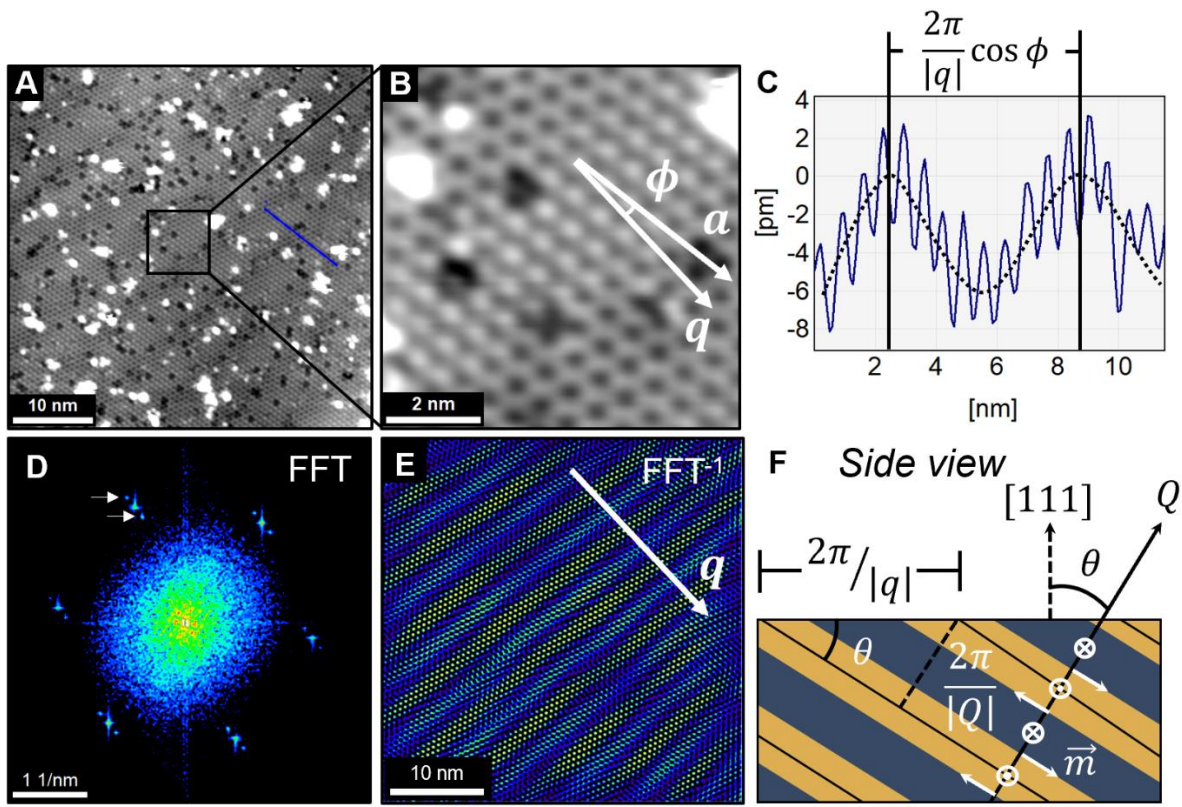

**Fig. 1 Atomic resolution SP-STM imaging of spin helices in MnGe. (A)** Topographic image of the MnGe(111) surface (-0.17 V, 0.54 nA). The helical magnetic texture is imaged as modulation in the atomic corrugation. **(B)** Zoomed view of the outlined area in **(A)**. Bright or dark lattice spots are observed depending on the relative alignment between tip and surface spins. In this area, the surface projection of the helical wave vector **q** is rotated from the atomic lattice vector by an angle $\phi \approx 14°$. **(C)** Line profile taken at the blue line in **(A)** along nearest neighbor atoms on the surface showing the helical periodicity of 5.96 nm. The dotted line is a guide to the eye. **(D)** FFT image of the area in **(A)**. Arrows indicate satellite peaks due to the surface projected **q**. **(E)** Inverse FFT image produced by passing only the atomic lattice and satellite peaks in **(D)**, allowing an unobstructed view of the stripe pattern and its effect on the atomic corrugation. **(F)** Schematic side view of the three-dimensional helical texture showing the relation between **Q**, **q** and the real-space modulations.

layers determines the structural and magnetic chiralities (*19*, *20*). The surface lattice constant from these images is 0.67 ± 0.01 nm, within experimental uncertainty of the expected bulk value for a (111) surface (0.678 nm). A low density of point defects on the surface are imaged with bright and dark contrast, but do not significantly affect the magnetic textures reported here.

In addition to the topographic information, Figures 1a-b also show a subtle (~ 5 pm) periodic modulation of the atomic corrugation reflecting the surface magnetic texture picked up by the SP-STM tip. This modulation is evident in Fig. 1b as alternating stripes of bright and dark triangular lattices, and in the topographic linecut shown in Fig. 1c. To better isolate the stripe pattern from the topography in Fig. 1a, we performed a Fast Fourier Transform (FFT) as shown in Fig. 1d. In addition to the primary hexagonal spot pattern associated with the MnGe atomic lattice, there are satellite spots corresponding to scattering vectors of $\pm q$, rotated by $\phi \sim 14°$ with respect to the lattice. An inverse FFT image of the area, computed with only the atomic lattice and satellite spots, clearly resolves the stripe pattern while removing obscuration from the point defects (Fig. 1e). In bulk MnGe crystals, a 3D hedgehog lattice was observed with Lorentz TEM (*12*), which would yield a 2D lattice projected onto the surface, in contrast to the observed stripe pattern here (see Supporting Information). Furthermore, from the FFT analysis we measure a stripe period of 5.96 nm, which is considerably larger than the helical pitch length for bulk MnGe (2.8 nm).

This stripe pattern is consistent with a 1***Q*** helical state in these MnGe thin films, in contrast to the 3***Q*** state observed in bulk crystals. *A priori*, one could explain the stripe contrast with helices anchored to the surface plane as reported for FeGe (*18*), but this is contradicted by the larger observed pitch length. Prior neutron scattering studies in MnGe thin films indicate that the magnitude of $Q = 2.2$ nm$^{-1}$ is unchanged from the bulk value (*17*), and we expect that any additional surface-specific effects, such as reduced exchange or surface DMI, would lead to an even smaller pitch length, in contrast to the observation. Instead, we consider the tilting of ***Q*** toward the film normal [111] direction by a polar angle $\theta$ which was invoked in the neutron studies (*17*). While a helix is described by a director $\overleftrightarrow{Q} = \pm \boldsymbol{Q}$, for simplicity we will use a wave vector ***Q*** and choose a positive projection along the surface normal $\hat{\boldsymbol{z}} = (111)$. We define the surface wave vector $q$ as the projection of $Q$ in the plane of the surface (i.e., q = Q sin$\theta$), so that the polar tilt angle can be related to the observed real-space periodicity by $\theta = sin^{-1}\left(\frac{2\pi}{Q \cdot 5.96 nm}\right) = 28.6°$, compared to the bulk angle of 54.7° with $Q$'s along (100). Our estimated $\theta$ is roughly consistent with the neutron studies, where a linearly decreasing tilt angle with decreasing film thickness down to 160 nm was attributed to strain-dependent magnetic anisotropy (*17*).

To directly probe the sensitivity of the spin helices to strain in real-space, we imaged regions of the film where small curvature is indicative of inhomogeneous strain. For example, in a different microscopic region of the sample shown in Figure 2a, three terraces were observed, separated in height by 1QL steps in the layered MnGe structure. Focusing on the middle terrace, topography line profiles show small (< 0.1%) but significant bowing and curvature of the surface along the horizontal direction (blue profile in Fig. 2b). We note that these images are atomically-resolved and the lattice spacing does not show any significant variation, but our experimental uncertainty (~ 0.5%) is larger than the 0.1% height variation shown in the image.

To examine the stripe patterns over larger distances in such areas, we simultaneously mapped the dI/dV signal which relates to the spin-dependent density of states. Figure 2c shows one such spatial map where faint magnetic contrast reveals stripes along different directions and more complicated patterns. During repeated imaging of this area, we observed a complete reversal in magnetic contrast associated with an inversion of the tip's spin polarization (Fig. 2d). Because the stripes appear with uniform contrast regardless of stripe direction, we infer that the tip spin polarization is perpendicular to the plane in both images. To confirm these stripes as magnetic in origin, we compute a difference image in Fig. 2e, as topographic or electronic contributions to the STM image would not invert under otherwise identical tunneling conditions. Instead of the straight stripes observed in Fig. 1, the stripe pattern in this area reveals intersections, bowing and terminations. We attribute these spin textures to local variations in the orientation of $Q$. For example, the observed stripe periodicity in Fig. 2e varies in the range of 6-10 nm depending on position, corresponding to respective variations $28° > \theta > 17°$ in the polar angle of $\mathbf{Q}$. The stripe curvature and intersection points however, indicate helical domains with distinct *azimuthal* angles of $\mathbf{Q}$.

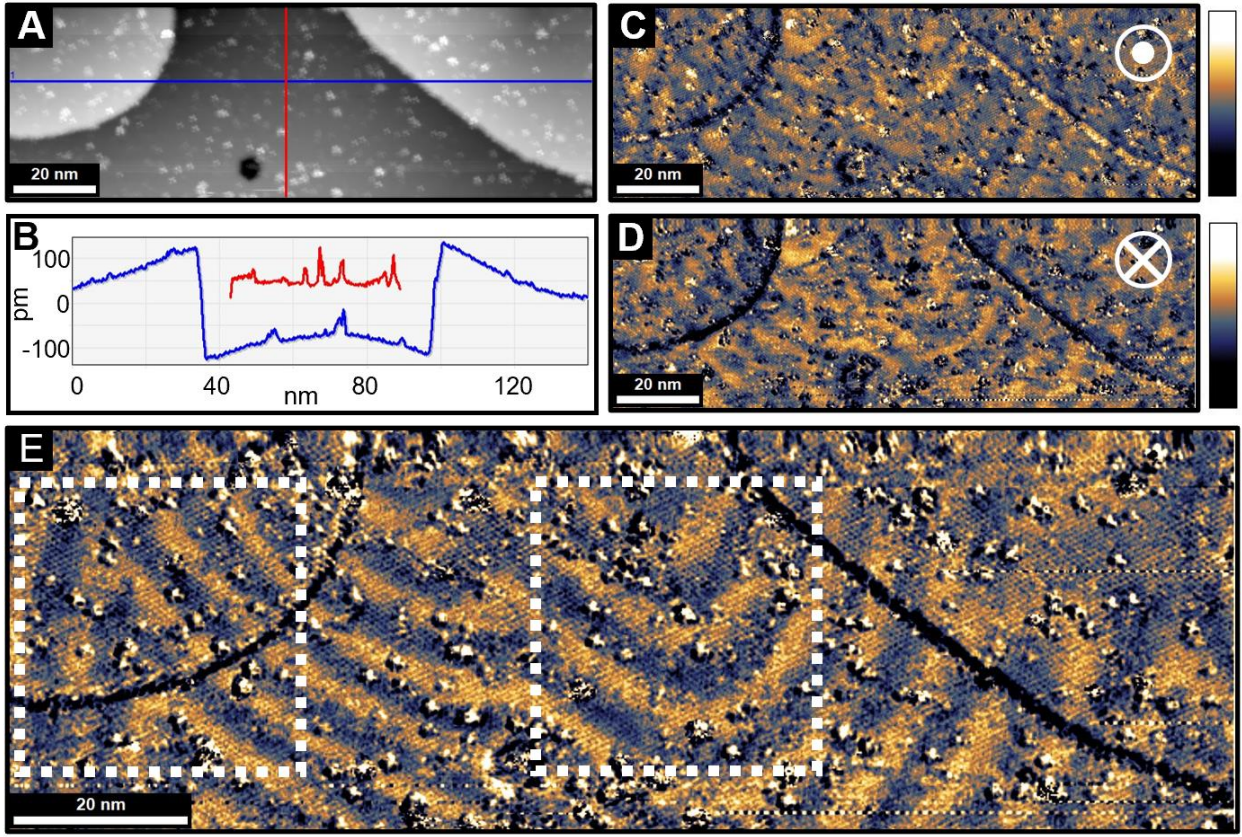

**Fig. 2 Spectroscopic imaging of helical domains in a bowed region of the surface. (A)** Topographic image of a bowed region of the surface containing three atomic terraces (0.17 V, 1.0 nA). **(B)** Line profile taken along the x (y) direction at the red (blue) line in **(A)**. **(C)-(D)** Subsequent dI/dV images of the same region as in **(A)** (0.17 V, 1.0 nA). The contrast of the helical texture is subtly visible and inverts upon reversal of the tip magnetization vector, which is represented by the symbol in the top right. **(E)** Difference image (C-D) clearly showing a variety of helical textures. Dotted lines indicate regions in Fig. 3.

To understand these features, we use a phenomenological model that builds on recent advances in the theory of topological domain walls in helimagnets (*18*) and uses inputs from neutron data (*17*) to constrain the magnetic anisotropy and hence the orientations of the helical wavevectors (see Supporting Information for more discussion). The structure of a domain wall between two helical regions depends primarily on the angle $\theta_{12}$ between their wavevectors $\mathbf{Q}_1$ and $\mathbf{Q}_2$. Three fundamental types of helical domain walls have been reported recently in MFM imaging of B20 FeGe (*18*). For $\theta_{12} \lesssim 85°$, one finds 'type I' walls, which are smooth, free of disclination defects or phase mismatch. We find that for the parameters relevant for our MnGe films, the type I domains walls are energetically preferred. Our micromagnetic modeling finds a distortion of the helices near the domain wall depending on whether the $Q_{1,2}$ vectors are oriented toward or away from the domain wall plane. We performed micromagnetic modeling of such a domain wall, as shown in Figs. 3a,c, and find two surface projections, depending on whether the $Q_{1,2}$ vectors are oriented toward or away from the intersection domain wall plane. For $Q$'s oriented toward the wall, the intersection plane is characterized by series of sharp, nested vertices along the domain wall (Fig. 3a). In contrast, for $Q$'s oriented away from the wall, the domain wall is characterized by a nesting of more gradual, curved helices (Fig. 3c). In both cases, the surface projections $q_1$, $q_2$ make an in-plane angle $\phi_{12} = 120°$ independent of $\theta_{12}$ (see Figs. 3a,c). The difference between these projections reflects rounding of the helical stripes in proximity to the domain wall and the surface.

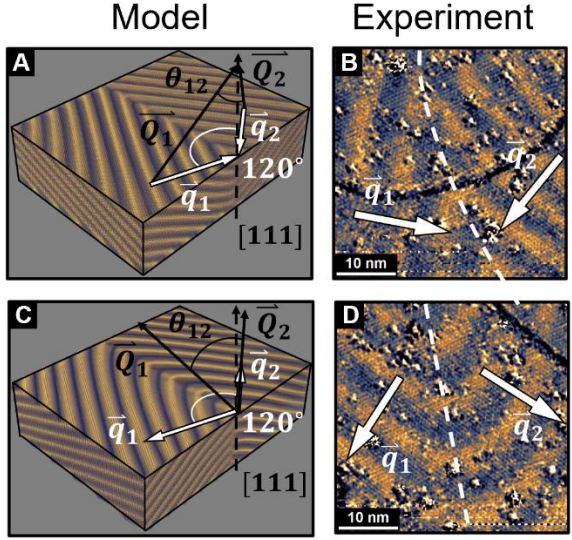

**Fig. 3 Comparison of model and observed helical domain walls (A)** Micromagnetic model of a domain wall where $Q_1, Q_2$ are separated by an angle $\theta_{12}$ and point toward the intersection plane. The surface projection shows a nesting of sharp vertices. **(B)** SPSTM image of this domain wall, with surface $q$'s denoted by the arrows. **(C)** Micromagnetic model of a domain wall where $Q_1, Q_2$ point away from the intersection plane, resulting in a nesting of more rounded helices. **(D)** SPSTM image of this domain wall, with surface q's denoted by the arrows. (B,D) are higher magnification views from the image in Fig. 2E.

In good agreement with our modeling, Figure 3 shows higher magnification SP-STM images of both projections of type I domain walls, extracted from regions in Fig. 2e. The domain wall in Fig. 3c shows the nested, sharp vertex-like structure expected from Fig. 3a. By considering the three dimensional nature of $Q_i$, we can extract the angle $\theta_{12}$ geometrically using Figs. 3a,c:

$$\cos\theta_{12} = \frac{Q_1 \cdot Q_2}{|Q_1||Q_2|} = \sin\theta_1 \sin\theta_2 \cos\phi_{12} + \cos\theta_1 \cos\theta_2.$$

Here $\theta_1 = 26°, \theta_2 = 19°$ are the polar angles calculated in each domain from the period of the stripes, and $\phi_{12} = 113°$ is estimated as the angle between the stripe patterns on either side of the domain wall. (The simplest model that ignores surface effects predicts a 120° angle, as noted above). This then gives an angle $\theta_{12} = 37°$ between $Q_1$ and $Q_2$ in this region, which is within the established regime for a type I domain wall (*18*). The SP-STM image in Fig. 3d shows the other surface projection of a type I domain wall, characterized by a nesting of rounded helical stripes, and can be analyzed in a similar way to give $\theta_{12} = 30°$, also within the type I regime.

More complex magnetic textures can be found at the intersections of domain walls. Our modeling shows that the intersection of two domain walls must necessarily involve at least one that is of type II or type III, which are energetically unfavorable. We find however, that three type I domain walls can meet along an axis perpendicular to the surface and lead to two distinct spin textures depending on the orientations of the $Q_i$'s. The spin texture in Fig. 4a results when all three $Q$'s are oriented toward or away from the intersection axis, and exhibits a core region that is wrapped with closed helical loops. These textures closely resemble topological defects known as 'target' states or $2\pi$-disclination defects (*21, 22*). The second 'π' texture, results from the arrangement of $Q_i$ as shown in Fig. 4b. As discussed in the Supplementary Information, both of these textures have nonzero topological charge density, concentrated in the vicinity of the domain walls, which oscillates in sign as one moves outward from the core of texture. In the absence of a well-defined boundary, however, neither the target nor the π-texture have a quantized topological charge. Our modeling also shows that the core of the target texture consists of a string of alternating hedgehogs and anti-hedgehogs that lies perpendicular to the surface.

Experimentally, we found both the target and π textures in a region of the film where there was nanoscale curvature in two dimensions. Figure 4c shows the conjunction of three type I walls to form the target texture, with closed helical loops wrapping around a central ~ 10 nm core. The isometric topographic image (Fig. 4e) and topographic linecuts (Fig. 4f) of this area indicate that the core is localized to the region of convex curvature to within a few nm in both the horizontal and vertical directions. Figure 4d shows the π texture in an adjacent region spanning two terraces separated by an atomic step across the middle. The local curvature in this region is slightly concave, and connects adjoining regions with target textures and convex curvature. Typically a handful of wrappings are observed around such textures before the pattern merges into a neighboring texture (see supporting information).

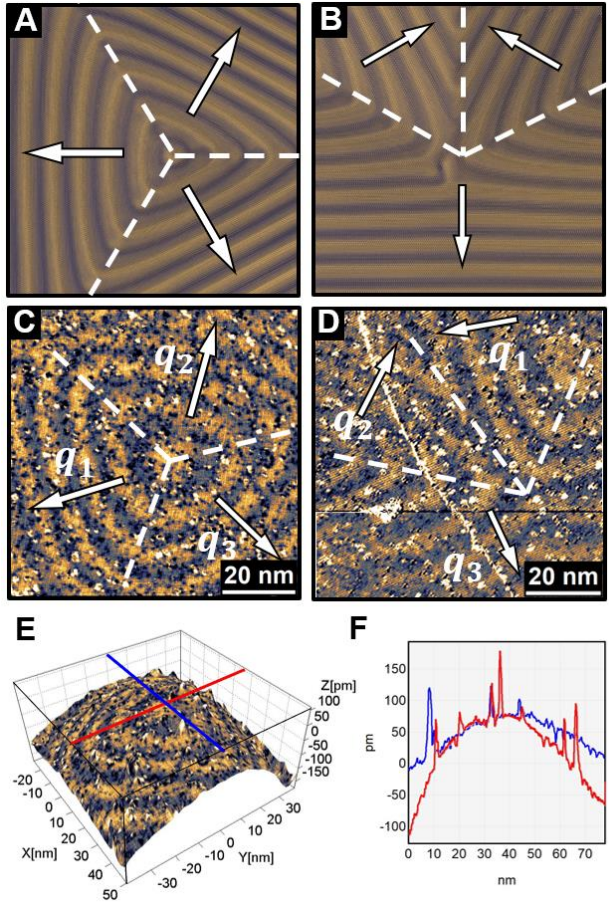

**Fig. 4 Modeling and observation of target and π textures (A)** Micromagnetic model of the target spin texture, with arrows indicating surface ***q***'s pointing away from the intersection axis. This results in rounded triangular rings wrapping a central core. **(B)** Model of the π texture where two q-vectors point inward toward each other and the third points away, resulting in two rounded domain walls meeting a sharp domain wall. **(C)** SP-STM image of the target texture (-0.31 V, 0.22 nA). **(D)** SP-STM image of a π texture (-0.31 V, 0.20 nA) **(E)** Three-dimensional view of the area hosting the target texture showing curvature of the surface. Topographic information is shown along the z-axis and the color scale is a dI/dV overlay from the image shown in **(C)**. **(F)** Line profiles taken along the x (blue) and y (red) directions to show in more detail the curvature of the surface in this area.

Applications of magnetic Skyrmions rely on the ability to manipulate these spin textures, which has been demonstrated with stimuli including STM pulsing (*23*), shaped current pulse densities (*24*) and optical excitation (*25*). We find the target texture can be similarly manipulated by local current/voltage pulses. Figures 5a-d show a sequence of SP-STM images where current/voltage pulses were applied to the core region using the STM tip. The initial state of the target texture in Fig. 5a has a bright core and several surrounding closed loops. After imaging, the STM tip was positioned over the core and a short (0.5 s) current/voltage pulse was applied. In addition to a shift of the core region by ~ 5 nm, we find a disclination defect (branch point) appeared in the subsequent SP-STM image (red circle in Fig. 5b), which represents a discrete change in the topological charge density within this region. Subsequent pulsing shifted the disclination defect closer to the core, which itself changes from bright to dark contrast (Fig. 5c). A final cycle of pulsing annihilated the disclination defect, and further shifted the core (Fig. 5d). The polarity and topology of this state are identical to the starting state (Fig. 5a), indicating that the system has been reversibly manipulated through a landscape of metastable states.

We also observed hysteretic behavior of the target texture with applied out-of-plane magnetic field, as shown by the sequence of SP-STM images in Fig. 5e. Focusing now on the core region, image (5) in Fig. 5e at +1T shows dark contrast, compared to the immediately preceding image #4, Fig. 5d. Ramping the field down to 0T (#6) shows a slight expansion of the wrapping and increased dark contrast of the core region. The core and wrapping contrast inverts when the field is ramped to -1T (#7), and this inverted contrast is maintained with only slight changes in wrapping and intensity when the field is ramped back down to 0T (#8). Importantly, images #6 and #8 (both taken at 0T) are identical except for the inverted magnetic contrast indicating a clear hysteresis effect, and image #8 and #4 show identical magnetic and topographic contrast, indicating that the tip's atomic termination and spin polarization was maintained throughout the series (c.f., Supporting information for more detailed comparisons). The reversible manipulation of these structures with applied current and magnetic field is a promising first step toward future applications.

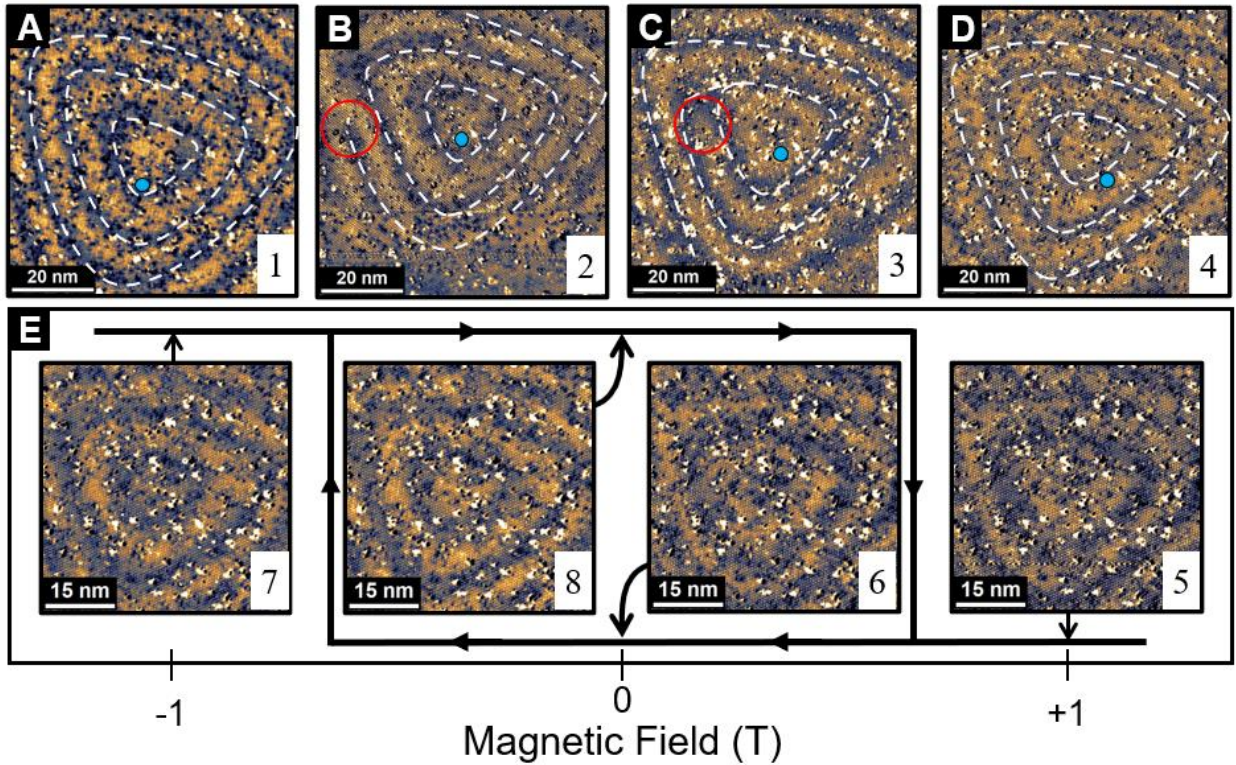

**Fig. 5 STM tip and magnetic field manipulation of a target texture (A)-(D)** SP-STM images of the target spin texture in different configurations. Between each image, current/voltage pulses are applied with the STM tip (~ 2.0 V, 0.5 sec, $I \lesssim 1\ \mu A$ due to 300 pm approach). White dotted lines are guides to the eye showing wrapping around the core region. The red circle indicates a disclination defect which is generated, moved and then annihilated. The blue dot represents the same atomically-registered fixed point in the images, showing motion of the core. **(E)** SP-STM images showing the magnetic field dependence of the texture. Image 5 is taken after image 4 with a +1T out of plane magnetic field, and shows reversed contrast. Subsequent imaging (6-8) was performed in the indicated loop, with a clear hysteresis effect evident upon comparison of images 6 and 8. All images are taken with (-0.3 V, 0.2 nA)

In summary, we have directly demonstrated that the magnetic textures in the unusual helimagnet MnGe are sensitive to local strain and surface effects, which accentuates the distinction between thin film and bulk materials. Micromagnetic modeling predicts a variety of topological domain wall structures depending on the orientation of the helical wavevectors, and these are confirmed experimentally through our SP-STM imaging. The core of the target texture consists of a string of alternating hedgehog and anti-hedgehog singularities, and it will be interesting to probe the connection of these features with the $3\boldsymbol{Q}$ hedgehog lattice phase to better understand the evolution of MnGe magnetism from thin film to bulk.

**Methods (further details available in supplement)**

**MBE growth**: A series of 10-500 nm thick MnGe(111) films were grown via molecular beam epitaxy on (P-doped, n-type) Si(111) substrates (*26*, *27*), which were treated with a buffered HF solution to remove the native oxide layer and H-terminate the surface. The freshly prepared Si(111) substrates were moved immediately into the UHV growth chamber to prevent re-oxidation, and annealed at ~800 °C until the 7x7 RHEED pattern characteristic of clean Si(111) emerged. The growth, which was monitored in real time using a 10 keV reflection high energy electron diffraction system (RHEED), was then started with a FeGe buffer layer after the Si substrate cooled down to 340 °C and followed by MnGe film at 300 °C. Both the FeGe buffer and MnGe film were grown using co-deposition from flux matched thermal effusion Fe, Mn and Ge cells. Shortly after growth, the MnGe films were transferred *in-vacuo* to a Ferrovac UHV suitcase with a base pressure of ~1.0 x$10^{-9}$ mbar and then to the SP-STM system.

**SP-STM measurements**: MnGe thin films were imaged without further surface preparation using a CreaTec low temperature (5 K) STM system equipped with a ±1 T, 2-axis vector magnet which allows for in-plane and out-of-plane magnetic fields. The SP-STM tips were etched from square bars of polycrystalline Cr, and cleaned by $Ar^+$ ion bombardment at a near coaxial angle. Tunneling spectra and spectral maps were obtained by adding a 20-50 mV modulation voltage at 1063 Hz to the DC bias and measuring the resultant change in current, dI/dV, using a lock-in amplifier. Image processing and analysis were performed using WxSM and SPIP.

**Modeling and micromagnetic simulations**: Our results are based on a simple phenomenological model, that builds on the recent theory of domain walls (DWs) in helimagnets (*18*) and uses inputs from MnGe neutron data (*17*) to constrain the magnetic anisotropy. We use a lattice model with energy

$$\mathcal{H} = -\sum_{\langle i,j \rangle} J\, \boldsymbol{m}_i \cdot \boldsymbol{m}_j + \sum_{\langle i,j \rangle} D\, \hat{\boldsymbol{r}}_{ij} \cdot (\boldsymbol{m}_i \times \boldsymbol{m}_j) - K \sum_j m_{i,z}^2$$

where $\boldsymbol{m}_i$ is a unit vector and $\langle i,j \rangle$ represent the nearest-neighbor bonds between sites *i* and *j*. Here *J* is the exchange coupling, *D* is the bulk DMI in a B20 material, and *K* is the anisotropy, with the *z*-axis chosen along (111). We place the spins on a BCC lattice which introduces an implicit higher order anisotropy term favoring the alignment of helical $Q$ vectors along [100] directions, when K=0. The value of K is chosen to move the orientation of the $Q$'s toward (111). The simulations we show are for $J = 1.0$, $D/J = 0.628$ and $KJ/D^2 = -0.14$. To understand the structure of domain walls (DWs) and their intersections, we chose initial configurations with specific textures, and relaxed the energy. To compare with SP-STM data, we look at the surface manifestation of textures whose energetics is governed by $\mathcal{H}$.

**References Cited**

1. A. Fert, N. Reyren, V. Cros, Magnetic skyrmions: advances in physics and potential applications. *Nat. Rev. Mater.* **2** (2017), doi:10.1038/natrevmats.2017.31.

2. A. Soumyanarayanan, N. Reyren, A. Fert, C. Panagopoulos, Emergent phenomena induced by spin-orbit coupling at surfaces and interfaces. *Nature*. **539**, 509–517 (2016).

3. R. Wiesendanger, Nanoscale magnetic skyrmions in metallic films and multilayers: a new twist for spintronics. *Nature Reviews Materials*. **1** (2016), doi:10.1038/natrevmats.2016.44.

4. C. H. Back, V. Cros, H. Ebert, K. Everschor-Sitte, A. Fert, M. Garst, T. Ma, S. Mankovsky, T. Monchesky, M. V. Mostovoy, N. Nagaosa, S. Parkin, C. Pfleiderer, N. Reyren, A. Rosch, Y. Taguchi, Y. Tokura, K. von Bergmann, J. Zang, The 2020 Skyrmionics Roadmap. *J. Phys. D: Appl. Phys.* (2020), doi:10.1088/1361-6463/ab8418.

5. K. M. Song, J.-S. Jeong, B. Pan, X. Zhang, J. Xia, S. Cha, T.-E. Park, K. Kim, S. Finizio, J. Raabe, J. Chang, Y. Zhou, W. Zhao, W. Kang, H. Ju, S. Woo, Skyrmion-based artificial synapses for neuromorphic computing. *Nature Electronics*. **3**, 148–155 (2020).

6. S. S. P. Parkin, M. Hayashi, L. Thomas, Magnetic Domain-Wall Racetrack Memory. *Science*. **320**, 190–194 (2008).

7. R. Tomasello, E. Martinez, R. Zivieri, L. Torres, M. Carpentieri, G. Finocchio, A strategy for the design of skyrmion racetrack memories. *Sci Rep*. **4**, 6784 (2014).

8. A. Fert, V. Cros, J. Sampaio, Skyrmions on the track. *Nature Nanotechnology*. **8**, 152–156 (2013).

9. S. Muehlbauer, B. Binz, F. Jonietz, C. Pfleiderer, A. Rosch, A. Neubauer, R. Georgii, P. Boeni, Skyrmion Lattice in a Chiral Magnet. *Science*. **323**, 915–919 (2009).

10. X. Z. Yu, N. Kanazawa, Y. Onose, K. Kimoto, W. Z. Zhang, S. Ishiwata, Y. Matsui, Y. Tokura, Near room-temperature formation of a skyrmion crystal in thin-films of the helimagnet FeGe. *Nat. Mater.* **10**, 106–109 (2011).

11. N. Kanazawa, J.-H. Kim, D. S. Inosov, J. S. White, N. Egetenmeyer, J. L. Gavilano, S. Ishiwata, Y. Onose, T. Arima, B. Keimer, Y. Tokura, Possible skyrmion-lattice ground state in the B20 chiral-lattice magnet MnGe as seen via small-angle neutron scattering. *Phys. Rev. B*. **86**, 134425 (2012).

12. T. Tanigaki, K. Shibata, N. Kanazawa, X. Yu, Y. Onose, H. S. Park, D. Shindo, Y. Tokura, Real-Space Observation of Short-Period Cubic Lattice of Skyrmions in MnGe. *Nano Letters*. **15**, 5438–5442 (2015).

13. M. Bornemann, S. Grytsiuk, P. F. Baumeister, M. dos S. Dias, R. Zeller, S. Lounis, S. Blügel, Complex magnetism of B20-MnGe: from spin-spirals, hedgehogs to monopoles. *J. Phys.: Condens. Matter*. **31**, 485801 (2019).

14. T. T. J. Mutter, A. O. Leonov, K. Inoue, Skyrmion instabilities and distorted spiral states in a frustrated chiral magnet. *Phys. Rev. B*. **100**, 060407 (2019).

15. Y. Fujishiro, N. Kanazawa, Y. Tokura, Engineering skyrmions and emergent monopoles in topological spin crystals. *Appl. Phys. Lett.* **116**, 090501 (2020).

16. M. Deutsch, P. Bonville, A. V. Tsvyashchenko, L. N. Fomicheva, F. Porcher, F. Damay, S. Petit, I. Mirebeau, Stress-induced magnetic textures and fluctuating chiral phase in MnGe chiral magnet. *Phys. Rev. B*. **90**, 144401 (2014).

17. N. Kanazawa, J. S. White, H. M. Rønnow, C. D. Dewhurst, D. Morikawa, K. Shibata, T. Arima, F. Kagawa, A. Tsukazaki, Y. Kozuka, M. Ichikawa, M. Kawasaki, Y. Tokura, Topological spin-hedgehog crystals of a chiral magnet as engineered with magnetic anisotropy. *Phys. Rev. B*. **96**, 220414 (2017).

18. P. Schoenherr, J. Müller, L. Köhler, A. Rosch, N. Kanazawa, Y. Tokura, M. Garst, D. Meier, Topological domain walls in helimagnets. *Nature Physics*. **14**, 465–468 (2018).

19. J. P. Corbett, T. Zhu, A. S. Ahmed, S. J. Tjung, J. J. Repicky, T. Takeuchi, J. Guerrero-Sanchez, N. Takeuchi, R. K. Kawakami, J. A. Gupta, Determining Surface Terminations and Chirality of Noncentrosymmetric FeGe Thin Films via Scanning Tunneling Microscopy. *ACS Appl. Mater. Interfaces*. **12**, 9896–9901 (2020).

20. S. V. Grigoriev, N. M. Potapova, S.-A. Siegfried, V. A. Dyadkin, E. V. Moskvin, V. Dmitriev, D. Menzel, C. D. Dewhurst, D. Chernyshov, R. A. Sadykov, L. N. Fomicheva, A. V. Tsvyashchenko, Chiral Properties of Structure and Magnetism in $Mn_{1-x}Fe_xGe$ Compounds: When the Left and the Right are Fighting, Who Wins? *Phys. Rev. Lett.* **110**, 207201 (2013).

21. F. Zheng, H. Li, S. Wang, D. Song, C. Jin, W. Wei, A. Kovács, J. Zang, M. Tian, Y. Zhang, H. Du, R. E. Dunin-Borkowski, Direct Imaging of a Zero-Field Target Skyrmion and Its Polarity Switch in a Chiral Magnetic Nanodisk. *Physical Review Letters*. **119** (2017), doi:10.1103/PhysRevLett.119.197205.

22. D. Cortés-Ortuño, N. Romming, M. Beg, K. von Bergmann, A. Kubetzka, O. Hovorka, H. Fangohr, R. Wiesendanger, Nanoscale magnetic skyrmions and target states in confined geometries. *Phys. Rev. B*. **99**, 214408 (2019).

23. N. Romming, C. Hanneken, M. Menzel, J. E. Bickel, B. Wolter, K. von Bergmann, A. Kubetzka, R. Wiesendanger, Writing and Deleting Single Magnetic Skyrmions. *Science*. **341**, 636–639 (2013).

24. W. Jiang, P. Upadhyaya, W. Zhang, G. Yu, M. B. Jungfleisch, F. Y. Fradin, J. E. Pearson, Y. Tserkovnyak, K. L. Wang, O. Heinonen, S. G. E. te Velthuis, A. Hoffmann, Blowing magnetic skyrmion bubbles. *Science*. **349**, 283–286 (2015).

25. N. Ogawa, S. Seki, Y. Tokura, Ultrafast optical excitation of magnetic skyrmions. *Sci Rep*. **5**, 9552 (2015).

26. A. S. Ahmed, B. D. Esser, J. Rowland, D. W. McComb, R. K. Kawakami, Molecular beam epitaxy growth of [CrGe/MnGe/FeGe] superlattices: Toward artificial B20 skyrmion materials with tunable interactions. *J. Cryst. Growth*. **467**, 38–46 (2017).

27. J. Engelke, D. Menzel, V. A. Dyadkin, Thin film MnGe grown on Si(111). *J. Phys.: Condens. Matter*. **25**, 472201 (2013).

28. T. Nattermann and V. L. Pokrovsky. Topological Defects in Helical Magnets. *Journal of Experimental and Theoretical Physics* 127, 922–32 (2018).

29. A. Bauer, A. Chacon, M. Wagner, M. Halder, R. Georgii, A. Rosch, C. Pfleiderer, and M. Garst. Symmetry Breaking, Slow Relaxation Dynamics, and Topological Defects at the Field-Induced Helix Reorientation in MnSi., *Physical Review B* **95**, 024429 (2017).

**Acknowledgments:** We thank Sara Mueller for helpful discussions on the data analysis **Funding:** Primary support was provided by DARPA Grant No. D18AP00008. J.G.S. and N.T. thank DGAPA-UNAM project IN101019, and CONACyT grant A1-S-9070 of the Call of Proposals for Basic Scientific Research 2017-2018 for partial financial support. Calculations were performed in the DGCTIC-UNAM Supercomputing Center, project LANCAD-UNAM-DGTIC-368. **Author contributions:** J.R. and J.P.C. performed SP-STM experiments and analyzed data, T.L.,T.Z. and A.S.A. grew the thin films, P.W. and M.R. performed theoretical modeling, N.T. and J.G.S. performed DFT calculations, and R.K., M.R. and J.A.G. helped in writing the manuscript and analyzing the results. **Competing interests:** Authors declare no competing interests. **Data and materials availability:** All published data and working files are available upon request.

# Supplementary Materials for

## Atomic scale visualization of topological spin textures in the chiral magnet MnGe

Jacob Repicky, Po-Kuan Wu, Tao Liu, Joseph Corbett, Tiancong Zhu, Adam Ahmed, N. Takeuchi, J. Guerrero-Sanchez, Mohit Randeria, Roland Kawakami, and Jay A. Gupta*
*Correspondence to: Gupta.208@osu.edu

**This PDF file includes:**

    Materials and Methods
    Supplementary Text
    Figs. S1 to S14

**Materials and Methods**

**MBE growth**: A series of 10-500 nm thick MnGe(111) films were grown via molecular beam epitaxy on (P-doped, n-type) Si(111) substrates (*26*, *27*), which were treated with a buffered HF solution to remove the native oxide layer and H-terminate the surface. The freshly prepared Si(111) substrates were moved immediately into the UHV growth chamber to prevent re-oxidation, and annealed at ~800 °C until the 7x7 RHEED pattern characteristic of clean Si(111) emerged. The growth, which was monitored in real time using a 10 keV reflection high energy electron diffraction system (RHEED), was then started with a FeGe buffer layer after the Si substrate cooled down to 340 °C and followed by MnGe film at 300 °C. Both the FeGe buffer and MnGe film were grown using co-deposition from flux matched thermal effusion Fe, Mn and Ge cells. Shortly after growth, the MnGe films were transferred *in-vacuo* to a Ferrovac UHV suitcase with a base pressure of ~1.0 x$10^{-9}$ mbar and then to the SP-STM system.

**SP-STM measurements**: MnGe thin films were imaged without further surface preparation using a CreaTec low temperature (5 K) STM system equipped with a ±1 T, 2-axis vector magnet which allows for in-plane and out-of-plane magnetic fields. The SP-STM tips were etched from square bars of polycrystalline Cr, and cleaned by $Ar^+$ ion bombardment at a near coaxial angle. Tunneling spectra and spectral maps were obtained by adding a 20-50 mV modulation voltage at 1063 Hz to the DC bias and measuring the resultant change in current, dI/dV, using a lock-in amplifier. Image processing and analysis were performed using WxSM and SPIP.

**Modeling and micromagnetic simulations**: Our results are based on a simple phenomenological model, that builds on the recent theory of domain walls (DWs) in helimagnets (*18*) and uses inputs from MnGe neutron data (*17*) to constrain the magnetic anisotropy. We use a lattice model with energy

$$\mathcal{H} = -\sum_{\langle i,j \rangle} J\, \boldsymbol{m}_i \cdot \boldsymbol{m}_j + \sum_{\langle i,j \rangle} D\, \hat{\boldsymbol{r}}_{ij} \cdot (\boldsymbol{m}_i \times \boldsymbol{m}_j) - K\sum_j m_{i,z}^2$$

where $\boldsymbol{m}_i$ is a unit vector and $\langle i,j \rangle$ represent the nearest-neighbor bonds between sites $i$ and $j$. Here $J$ is the exchange coupling, $D$ is the bulk DMI in a B20 material, and $K$ is the anisotropy, with the $z$-axis chosen along (111). We place the spins on a BCC lattice which introduces an implicit higher order anisotropy term favoring the alignment of helical $\boldsymbol{Q}$ vectors along [100] directions, when K=0. The value of K is chosen to move the orientation of the $\boldsymbol{Q}$'s toward (111). The simulations we show are for $J = 1.0$, $D/J = 0.628$ and $KJ/D^2 = -0.14$. To understand the structure of domain walls (DWs) and their intersections, we chose initial configurations with specific textures, and relaxed the energy. To compare with SP-STM data, we look at the surface manifestation of textures whose energetics is governed by $\mathcal{H}$.

**Supplementary Text**

**Theoretical modeling and micromagnetic simulations**

In this Supplement we describe the theoretical modeling of SP-STM data on the surface of a MnGe thin film grown by MBE on Si(111). We show that we can build on the recently developed theory of domain walls in helimagnets (*18, 28*) to obtain a consistent and detailed description of the SP-STM data. Our most important results are as follows.

(1) We show that the 2D wave vectors of the stripe domains probed by SP-STM are the surface projections of the 3D $\boldsymbol{Q}$'s that are consistent with neutron scattering (*17*) data on MnGe thin films.

(2) We show that the domain walls (DWs) between helical domains are smooth, i.e., free of disclinations or phase mismatch. The two domains on either side of a DW are characterized by 2D (in-plane) wavevectors with 120° angles between them.

(3) The only energetically favorable DW configurations involve either a single DW or three DWs. A single DW leads to a "vee"-like pattern of helices on the surface (see Fig. 5), which can be either sharp or rounded, depending on the orientation of the 3D $Q$'s with respect to the sample surface.

(4) The intersection of three DWs can occur in two district ways: a target-like texture (see Fig. 7(c)) or a π-texture (Fig. 7(c)), both of which have been seen by SP-STM. Both these textures have a non-zero topological charge density, which oscillates in space, and thus they do not have a well-defined total topological charge.

(5) The core of the target-texture has a string of hedgehogs and anti-hedgehogs in the direction perpendicular to the surface.

Let us begin by noting that the "3$Q$ hedgehog/anti-hedgehog crystal" seen in *bulk* MnGe samples by Lorentz TEM (*12*), and consistent with topological Hall data (*11*), is quite different from the helical and skyrmion crystal phases seen in *all* other B20 materials. The theoretical understanding of why a 3Q hedgehog crystal is stabilized in bulk MnGe is not well established.

For our purposes here, it is sufficient to note that *if* we assume that the $H = 0$ ground state in (111) MnGe thin films is a 3Q hedgehog crystal, then SP-STM should see the "triangular lattice" spin textures plotted in Fig. 1. (The three panels correspond to different choices of the angles that the three $Q$'s make with the surface normal). It is evident, however, that our SP-STM data on the (111) surface is qualitatively inconsistent with such a state.

We will show that all of our SP-STM data can be understood in terms of helical domains and their domain walls (DWs), rather than the 3$Q$ state, as described in detail below. The structure of DWs in helimagnets, which has been elucidated only very recently (*18*), is much more complex than in simple ferromagnets. This stems from the non-trivial topology of the order parameter space $SO(3)/Z_2$ of an isotropic helimagnet (*28*) since a helix is described by a director $\overleftrightarrow{Q}$ rather than a vector. Our analysis below builds on recent advances in the theory of topological DWs in helimagnets (*18, 28*).

Let us define our notation at the outset. We often find it simpler to describe a helix by a 3D wave vector $\boldsymbol{Q}$ rather than a director. Its magnitude is given by $2\pi$ over the helical pitch. However, we must pick a convention for its direction since $\overleftrightarrow{Q} = \pm \boldsymbol{Q}$ describe the *same* helix. We choose a convention where $\boldsymbol{Q}$ has positive projection along the surface normal $\hat{\boldsymbol{z}} = (111)$. This is permissible because all $\boldsymbol{Q}$'s of interest in our system will turn out to have a non-zero component along $\hat{\boldsymbol{z}}$, as we show next. We will use lower case $\boldsymbol{q}$, which lies in the sample surface, to denote the 2D projection of the 3D wave vector $\boldsymbol{Q}$ onto the surface; see Fig. 2.

The SP-STM rule out the possibility of the helical $\overleftrightarrow{Q}$'s lying in the plane of the surface. First, there is a *quantitative* problem if the $\overleftrightarrow{Q}$'s were to lie on the surface. The surface helical period $\simeq$ 10 nm seen by SP-STM is larger than the bulk period $\simeq$ 3nm seen by neutron scattering (*17*). The surface could have an effectively reduced exchange $J$ or an enhanced DMI $D$ (due to interface DMI in addition to the bulk DMI in the B20 material), but both these effects would lead to a decrease in the helical period $2\pi J/D$ relative to the bulk, just the opposite of what is observed.

Second, assuming in-plane $\overleftrightarrow{Q}$'s for MnGe leads to a *qualitative* failure in accounting for the structure of the DWs. SP-STM sees that the in-plane directors of the domains on either side of a DW make an angle of 120°. The theory of helimagnetic DWs (*18*) unambiguously predicts that, for such $\overleftrightarrow{Q}$'s lying entirely in-plane, the stable DW with the lowest energy should be a "zig-zag" wall with an array of π-disclinations, as discussed in detail below. This is totally inconsistent with DWs seen by SP-STM in MnGe.

We show that both these problems are resolved when one considers helical domains with $\boldsymbol{Q}$'s that have a substantial component along $\hat{\boldsymbol{z}} = (111)$. We have independent information about the $\boldsymbol{Q}$'s from neutron scattering experiments of Kanazawa *et al.* (*17*) which show six peaks at $\pm \boldsymbol{Q_i}$ (i = 1,2,3) each of which is tilted at an angle θ with respect to the surface normal $\hat{\boldsymbol{z}}$. The neutron data (*17*) shows that the angle θ is a decreasing function of the sample thickness and, linearly extrapolating to our film thickness of 80 nm, we estimate $\theta \simeq 15° \pm 5°$. On the other hand, we can also estimate $\theta$ using $\sin\theta = q/Q$, where $Q$ is the magnitude of the 3D wave vector obtained from neutrons and $2\pi/q$ is the average pitch of a stripe domain measured by SP-STM. Using $Q \simeq 2.2$ nm$^{-1}$ seen by neutron data (independent of thickness over a broad range 735 nm down to 160 nm), and $2\pi/q = 5.96$ nm as measured by SP-STM, we find $\theta = 28.6°$. We note that SP-STM finds a range of $q$-values at different locations on the surface, and this leads to a corresponding range of estimates of $\theta \simeq 17° - 29°$. We thus find rough consistency between these $\theta$ estimates, which is reasonable considering that this angle is highly sensitive to the magnetic anisotropy in the thin film, and that our film thickness of 80 nm is a factor of 2 smaller than the thinnest samples measured by neutrons.

The $\boldsymbol{Q}$-vectors we are using to model our thin film data in terms of single-$\boldsymbol{Q}$ helical domains, would also lead to six spots $\pm \boldsymbol{Q_i}$ (i = 1,2,3) in a neutron experiment, and could not be distinguished in a simple way from the scattering intensity arising from a 3$\boldsymbol{Q}$ structure, which occurs in bulk single crystals. As already emphasized, our real space imaging shows that we indeed have single-$\boldsymbol{Q}$ helical domains in our 80 nm thin film.

In the next section we summarize the micromagnetic modeling of helical domains. We then turn to the surface manifestation of DWs between helical states with 3D wavevectors $\pm \boldsymbol{Q_i}$ (i = 1,2,3) and compare it with our SP-STM data in detail starting in Section II.

**I. Micromagnetic modeling of helical state of MnGe**

Our goal is to see how far we can use from *bulk* energetic considerations to explain the structure and stability of helical domains and their DWs seen on the *surface*. Toward this end, we describe the spin textures in MnGe using the energy density

$$\mathcal{H} = J(\nabla \boldsymbol{M})^2 + D\boldsymbol{M} \cdot (\nabla \times \boldsymbol{M}) - K(\boldsymbol{M} \cdot \hat{\boldsymbol{n}})^2 + V_4(\boldsymbol{M}) + W_6(\boldsymbol{M}) \tag{1}$$

with $|\boldsymbol{M}(\boldsymbol{r})| = 1$. The first term in eq. (1) describes ferromagnetic exchange of strength $J$ and the second term is the Dzyaloshinskii-Moriya interaction (DMI) in a B20 crystal with strength $D$. The third term describes the uniaxial anisotropy and we choose $\hat{\boldsymbol{n}}$ to be the surface normal $\hat{\boldsymbol{z}} = (111)$. $V$ and $W$ are 4th and 6th order anisotropy terms (respectively) that will be described below. We have set the external field $\boldsymbol{H} = 0$.

The symmetry-allowed anisotropy terms in eq. (1) are chosen to stabilize the observed helical textures. From neutron scattering data (*17*) we see that, as the sample thickness decreases, the directors of the helices move away from the crystalline axes (100) and toward the surface normal (111). In other words, the angle $\theta$ between $\widehat{\boldsymbol{Q}}_i$ and the surface normal $\hat{\boldsymbol{z}}$ (see Fig. 2) decreases with film thickness. There is, in addition, a small rotation of the directors about (111) that we will also comment on.

To understand the neutron observations, we include a weak magneto-crystalline anisotropy term

$$V_4(\boldsymbol{M}) = V(M_X^4 + M_Y^4 + M_Z^4)$$

Here we find it useful to introduce coordinates $\widehat{\boldsymbol{X}} = (100)$, $\widehat{\boldsymbol{Y}} = (010)$ and $\widehat{\boldsymbol{Z}} = (001)$ along the crystalline cubic axes. The corresponding components of the magnetization are written as $\boldsymbol{M} = (M_X, M_Y, M_Z)$. Note that upper case $\widehat{\boldsymbol{Z}} = (001)$ is distinct from the lower case $\hat{\boldsymbol{z}} = (111)$ used for the surface normal.

The uniaxial anisotropy $K$ is negligible in the bulk, and choosing $V < 0$ the lowest energy spin configurations correspond to helices with wave-vectors $\boldsymbol{Q}_1$, $\boldsymbol{Q}_2$, and $\boldsymbol{Q}_3$ along $\widehat{X} = (100)$, $\widehat{Y} = $

(010) and $\hat{Z} = (001)$. This corresponds to $\theta = \tan^{-1}(\sqrt{2}) = 54.7°$ in the bulk limit. With decreasing film thickness, the data imply an increasing *easy-plane* anisotropy $|K|$ with $K < 0$. This leads to the three $Q_i$'s tilting toward the surface normal $\hat{z} = (111)$ as the film thickness decreases, and thus a decrease in $\theta$. We note that the in-plane projections $q_i$'s always make $120°$ angles with each other independent of $\theta$ (see Fig. 2), which will play an important role below.

By minimizing the energy of helix as a function of its direction, we find that $Q$ moves from (100) towards the (111) with increasing $|K|$, but then $\theta$ jumps discontinuously from $15°$ to $0°$ at $|K/V_4| \simeq 1.18$. We note in passing that such a transition is consistent with previous mean field calculations (*29*) for MnSi, although their argument was based on an energy written as a function of $\overleftrightarrow{Q}$ rather than an anisotropy energy written in terms of $M$.

We always work in a regime where $\theta \simeq 28.6°$ as appropriate for our experiments. For our numerical micromagnetic simulations, we found it convenient to mimic the effect of a weak magneto-crystalline anisotropy by simply placing our spins on a BCC grid, and then varied the uniaxial anisotropy $K$ to obtain the desired $\theta$. (See Section VII for details on simulations).

Following ref. (*29*), symmetry permits a sixth-order anisotropy term which causes a rotation of $\overleftrightarrow{Q}_i$'s about $\hat{z} = (111)$ by an angle $\phi$. This $W_6$ anisotropy is necessary to account for the non-zero $\phi$ observed in both the neutron scattering (*17*) experiments as well as in our SP-STM images, where we see $\phi \simeq 14°$; see Fig. 1B of the main text. However, this term has little influence on for the important properties of the helical domains and their DWs that we are focusing on (other than the non-zero $\phi$). Therefore, to keep things simple, we choose to omit it in the subsequent discussion and in our numerical simulations.

We end this section by noting some differences between the helical phases in MnGe and FeGe. The 3 nm helical pitch in MnGe implies a much larger $D/J$ compared to FeGe, which has a 70 nm pitch. Their anisotropies also seem to be very different, since in FeGe the helical directors lie in the plane of the surface (*18*), while in MnGe they are at a non-zero angle $\theta$ to the surface normal. These differences will play a crucial role in understanding the differences between the textures seen on the surfaces of FeGe by MFM (*17*) and of MnGe by SP-STM (in this paper).

## II. Domain walls in MnGe

In this Section we first summarize the results of Schoenherr *et al.* (*18*), since our analysis builds on these results. We then show how the *three dimensional* $\boldsymbol{Q}$-vectors, consistent with neutron data (*17*), help us understand the surface textures observed by SP-STM

Consider a DW between two helical domains with directors $\vec{\vec{Q}}_i = \pm \boldsymbol{Q}_i$ and $\vec{\vec{Q}}_j = \pm \boldsymbol{Q}_j$. By choosing the convention that $\boldsymbol{Q}_i \cdot \boldsymbol{Q}_j \geq 0$, the angle between the directors can be described in terms of $\theta_{ij}^- = \angle \boldsymbol{Q}_i \boldsymbol{Q}_j$ and $\theta_{ij}^+ = \angle \boldsymbol{Q}_i(-\boldsymbol{Q}_j) = 180° - \theta_{ij}^-$, with $\theta_{ij}^- \leq 90° \leq \theta_{ij}^+$. When there is no ambiguity, we simply call this angle $\theta_{12}$ to simplify notation. We use $\alpha$ to denote the angle between a director and a domain wall; see Fig. 3(b).

Based on MFM experiments on FeGe and micromagnetic simulations Schoenherr *et al.* (*18*) found that stable DWs in helimagnets come in three types (I, II, III), depending on the angle $\theta_{12}$ between $\boldsymbol{Q}_1$ and $\boldsymbol{Q}_2$ characterizing the two domains on either side of the DW. (We see from Fig. 3(b) that for type-I DWs $\theta_{ij} = \theta_{ij}^-$, while for Type-III DWs $\theta_{ij} = \theta_{ij}^+$).

In Fig. 3 (a) and (b) we summarize of the main results of ref. (*18*). Panel (a) shows the DW, characterized by the angle $\alpha$, which is the minimum energy configuration for a given value of $\theta_{12}$. In panel (b) we show the three types of DWs: their helical stripes, $\boldsymbol{Q}$ vectors, $\theta_{12}$ and $\alpha$.

Type I DWs are stable (*18*) for $\theta_{12} \lesssim 85°$. They are characterized by the angle $\alpha = \theta_{12}/2$ between the domain wall and either wave vector. Thus a Type I DW lies along the bisector of the two wave vector.

Type II DWs are stable for $85° \lesssim \theta_{12} \lesssim 140°$. This DW has a "zig-zag" shape and is composed of a series of $\pm\pi$ disclinations (*28*). The angle $\alpha \simeq 90°$ due to the tendency that DW tends to be parallel to the stripes of one domain. We also note that for $\theta_{12} \simeq 85°$ or $\theta_{12} \simeq 140°$ a special type II DW is found that does not have a zig-zag shape, and the DW is simply parallel to the stripes of one domain as shown in Fig 6(d).

Type III DWs are stable for $\theta_{12} \gtrsim 140°$. Here too $\alpha = \theta_{12}/2$ and the DW lies along the bisector of two the wave vectors, like the Type I case. There is, however, either a π-phase shift between the helices in the two domains or a small distortion at the DW, unlike the Type I case where the helices are phase-matched and smooth across the DW.

We note that, for both type I and type III DWs, the DW must lie along the bisector of the two $Q$ vectors since the magnitude of projection of wave vectors $Q_1$ and $Q_2$ on the DW should be the same. If the DW were not to be the bisector of $Q_1$ and $\pm Q_2$, the changing phase shift across the DW would be energetically highly unfavorable (*18*).

Now we apply these results to understand the SP-STM data on MnGe. It is clear that the DWs observed by SP-STM (see, e.g., Fig. 3c) are *qualitatively* inconsistent with the assumption of $Q$-vectors lying in the plane of the sample surface. These DWs are oriented along the bisector of the 2D wave-vectors of two phase-matched helical domains. This seems like a Type I DW, but such a DW *cannot* be stable for *in-plane* wave-vectors that make an angle of 120°. For in-plane $Q$-vectors at 120°, the theory unambiguously predicts a Type II DW, in contradiction with what is not observed.

Next, we consider 3D $\boldsymbol{Q}$-vectors, consistent with neutron scattering (17), and show how these help us understand the surface textures observed by SP-STM. We write $\boldsymbol{Q}_1 = (q_{1x}, q_{1y}, |q_1|/\tan\theta)$ and $\boldsymbol{Q}_2 = (q_{2x}, q_{2y}, |q_2|/\tan\theta)$ with $|Q_1| = |Q_2|$. Note that each $\boldsymbol{Q}_i$ makes an angle $\theta$ with respect to the surface normal (111) that $\boldsymbol{q}_i = (q_{ix}, q_{iy})$ are the 2D are the 2D projections; see Fig. 2.

Consider the type-I DW, the angle between the 3D wavevectors $\theta_{12} = \theta_{12}^- = \angle \boldsymbol{Q}_1 \boldsymbol{Q}_2$ is given in terms of $\theta = \angle \boldsymbol{Q}_i \hat{\boldsymbol{z}}$ and the angle $\phi_{12} = \angle \boldsymbol{q}_1 \boldsymbol{q}_2 = 120°$ by the expression

$$\cos\theta_{12} = \frac{\boldsymbol{Q}_1 \cdot \boldsymbol{Q}_2}{|Q_1|^2} = \frac{|q_1||q_2|/\tan^2\theta + |q_1||q_2|\cos\phi_{12}}{|q_1|^2/\sin^2\theta} = \cos^2\theta - \frac{1}{2}\sin^2\theta. \quad (2)$$

Using $\theta = 28.6°$ we find that $\theta_{12} = 49.5°$, for which Type I DWs are stable in the bulk as explained above. On the (111) surface of MnGe, this DW will look smooth and phase-matched like a Type I DW but with an angle between $\phi_{12} = \angle \boldsymbol{q}_1 \boldsymbol{q}_2 = 120°$ between the 2D wave-vectors on the surface, exactly as seen by SP-STM. We will discuss in Section III below some details of the distortions that occur near DWs.

We next use Eq. (2) to get a deeper understanding of the DWs in MnGe. In Fig. 4, we plot $\theta_{12}$ (blue curve) as a function of the tilt-angle θ. We find that $\theta_{12} \approx \sqrt{3}\theta$ is a good approximation to the solution of Eq. (2) over the entire range $0 < \theta < 54.7°$ of interest.

We must remember that a helix described by a director $\overleftrightarrow{Q}$. When considering type I DW, we discussed the case $\theta_{12} = \theta_{12}^-$, but $\theta_{12} = \theta_{12}^+ = 180° - \theta_{12}^-$ is also a solution, as can be seen by changing the sign of either one the $\boldsymbol{Q}_1, \boldsymbol{Q}_2$ in Eq. (2). This second solution is plotted as the orange curve in Fig. 4. Using the results shown in Fig. 3(a), we mark in Fig 4 the regimes where Type I, II and III DWs are stabilized.

For the range of $\theta$ values of interest ($\theta \simeq 17 - 29°$) for our MnGe thin films, the $\theta_{12}^{-}$ solution (blue curve in Fig. 4) leads to a Type I DW, while the $\theta_{12}^{+}$ solution (orange curve) can lead to type II and type III. (For further discussion of Type II and III DWs see Fig. 6 and associated text below). We emphasize that Type I DWs, which are smooth and free of disclinations or phase mismatch, are greatly favored energetically over Type II and Type III DWs; see Fig. 3 of ref. (*18*). This is clearly consistent with the fact that essentially all the DWs seen by SP-STM in MnGe are type I DWs.

### III. Distortions near type I domain walls

In the SP-STM experiments we find that the helices become slightly distorted near Type I DWs, some of which look "rounded and smooth" (see Fig. 5(a)), while others are "sharp" (Fig. 5(b)). From our simulation we found that this behavior can be understood in terms of the minimum energy configuration depending on how the two $\boldsymbol{Q}$-vectors defining the DW splay with respect to the surface and the DW.

Recall that we defined our $\boldsymbol{Q}$-vectors such that $Q_z > 0$, where $\hat{\boldsymbol{z}} = (111)$ is the surface normal. We find that if $\boldsymbol{Q_1}$ and $\boldsymbol{Q_2}$ point away from the DW, the DW is rounded. In contrast, if $\boldsymbol{Q_1}$ and $\boldsymbol{Q_2}$ point towards the DW, the DW is sharp. These two cases are shown in Fig 5(a) and (b) respectively. The cross-sectional view of the spin textures near the DW shown is Fig. 5(c) is instructive. Note that the DW lies in a plane perpendicular to the figure. We see two interesting effects: the rounding of the helical stripes as they approach the DW in the bulk of the film, and the distortions of the stripes close to the surface. Both of these effects are connected to the different appearance of the type I DW on the surface depending on how the $\boldsymbol{Q}$-vectors splay. We should note that in the simplest phenomenological model we are analyzing, we have not taken into account any surface energy contributions.

**IV. Intersection of domain walls**

In this section we analyze states with *intersecting DWs* in our system. Our main conclusions are that: (i) the intersection between *two* DWs is energetically costly, and (ii) *three* DWs can intersect in only two ways, either the "target" structure or the $\pi$-texture (see Fig. 7(c)), both of which are observed in our SP-STM experiments.

Without loss of generality, we can discuss intersecting DWs that arise from domains with $\boldsymbol{Q}$-vectors are along the directions

$$\begin{cases} \boldsymbol{Q}_1 = (\sin\theta,\ 0,\ \cos\theta) \\ \boldsymbol{Q}_2 = \left(-\frac{1}{2}\sin\theta,\ \frac{\sqrt{3}}{2}\sin\theta,\ \cos\theta\right) \\ \boldsymbol{Q}_3 = \left(-\frac{1}{2}\sin\theta,\ -\frac{\sqrt{3}}{2}\sin\theta,\ \cos\theta\right). \end{cases} \quad (3)$$

Note that these each (normalized) $\boldsymbol{Q}$ is tilted at an angle $\theta$ to the surface normal (111), chosen as the z-axis, and the in-plane projections of these $\boldsymbol{Q}$ vectors make $120°$ angles with each other. We focus primarily on Type I DWs, which must lie along the bisector of $\boldsymbol{Q}_i$ and $\boldsymbol{Q}_j$ (see Fig. 2). We see from eq. (3) that the DW normal defined by $\hat{\boldsymbol{n}}_{ij} = \widehat{\boldsymbol{Q}}_i - \widehat{\boldsymbol{Q}}_j$ (see Fig. 3) has no z-component. Thus we see that the type I DW must be perpendicular to the sample surface (which is the *xy*-plane).

Flipping sign of a wave-vector to $-\boldsymbol{Q}_i$, leads to the second solution $\theta_{ij}^{\dagger}$ discussed above,

and leads to a Type III DW for $\theta \lesssim 23°$ and to a Type II DW for larger $\theta's$. We investigate this further, even though we discarded it earlier based on energetic considerations. Such a DW has a normal vector $\hat{\boldsymbol{n}}'_{ij} = \hat{\boldsymbol{Q}}_i + \hat{\boldsymbol{Q}}_j$. Unlike the normal $\hat{\boldsymbol{n}}_{ij}$ to a Type I DW, the Type III or Type II normal $\hat{\boldsymbol{n}}'_{ij}$ has large *z*-component (Type II DW is parallel to the strip of helices, so the angle between $\hat{\boldsymbol{n}}'_{ij}$ and *z*-axis is also $\theta$), so these domain walls are *not* perpendicular to the sample surface (*xy*-plane).

**Textures with two domain walls**

We now look at the intersection of two DWs. If we assume that one DW is of Type I (lowest energy), then the other DW must necessarily be of Type II or Type III, as seen from see Fig. 6 (a,b,c,d). The detailed structure of such two DW intersections is quite complex. It is, however, worth pointing a simple fact about type II and III DWs. The planes defined by these DWs in 3D are *not* perpendicular to the thin film surface, in contrast the type I DW that is always in a plane perpendicular to the surface. Type II and Type III DWs are energetically costly and they have never been seen in our SP-STM data. Thus we do not discuss the intersection of two DWs further.

**Textures with three domain walls**

We next argue that the intersection of three domain walls leads to either the "target texture" or the "$\pi$-texture" shown in Fig 7(c). Both are seen in experiments.

Recapping the discussion above: only Type I DWs exist in our system, and the normal vector to such a DW $\boldsymbol{n}_{ij} = \hat{\boldsymbol{Q}}_i - \hat{\boldsymbol{Q}}_j$ lies in the *xy*-plane (as is easy to see from Eq. 3). Thus *on the surface* the DW lies along the vector $\boldsymbol{v}_{ij} \perp \boldsymbol{n}_{ij}$, where the $\boldsymbol{v}_{ij}$'s also lie in the *xy*-plane; see Fig 7(a). These constraints make the projection of the texture to the *xy*-plane easier to visualize.

Now the three (projected) $q_i$'s are at 120° to each other, and so are the three DW $v_{ij}$'s, given that the Type I DW must be a bisector of the domain wave-vectors. This means that configurations such as the ones shown in Fig 7(b) are disallowed. Thus the textures shown in Fig 7(c) are *the only two allowed configurations with three intersecting DWs*. A simple way to understand why there are only two distinct ways in which three domains meet is to note that that there are only two distinct cyclic permutations of three wavevectors. The target texture, shown at the top of panel (c), is highlighted in the main text (see Fig. 4c of the paper). The $\pi$-texture shown at the bottom of panel (c) is also seen in the SP-STM data (see Fig 8).

## V. Topological charge density and core of target and π-textures

We show in Fig 9 (a,b) that the topological charge density in the target and $\pi$ textures is concentrated near each of the three DWs, which are also the regions where the helical stripes bend the most. Each "vertex" of the "triangular" target texture harbors a $\pm 1/3$ topological charge, so that each "ring" has a charge $\pm 1$, with the signs alternating as one moves outward from the core. The situation with the $\pi$ texture is very similar, with the topological charge concenttrated at the intersection of the three DWs with the helical stripes. As a result of the oscillation in sign as one moves outward from the core, there is no fixed total charge – or skyrmion number – for an isolated texture, i.e., one for which there no length-scale at which it terminates.

This situation is similar to the "circular target texture", or $2\pi$-disclination (*28*), considered in the supplementary material of ref. (*18*). Here the topological charge density $\rho(r)$ oscillates in sign as a function of $r$, but is uniformly spread over $2\pi$ angle with a $1/r$ decay, so that it is normalized on a ring. Here too the total topological charge $\int d\phi \int_0^R dr\, r\, \rho(r)$ is ill-defined since it depends on the cutoff $R$.

We conclude this section by discussing the core of the target and π textures. From the cross-sectional view shown in Fig. 10, we see that the core of the target texture consists of a string of hedgehogs and antihedgehogs along the along the $\hat{z} = (111)$ axis. This core structure is reminiscent of (or a is a 1D analog of) the $3Q$ configuration seen in bulk single crystals. However, we find the core of the π texture does not show these singularities.

## VI. Simulation details

To simulate the magnetic system, we use a magnetic lattice model with unit vectors $m$. To simulate a model described in Eq. (1), there are exchange interaction, bulk DMI and easy-plane anisotropy. In the bulk limit where $K=0$ the helical wave-vectors are along (100), (010), (001). Instead of adding the $V_4$ term in eq. (1) we introduce this tendency by using a body center cubic lattice (see Fig. 11), where this anisotropy arises implicitly from the lattice when the helical pitch length is not much larger than lattice spacing. Our model is

$$\mathcal{H} = \sum_{\langle i,j \rangle} J \boldsymbol{m}_i \cdot \boldsymbol{m}_j + \sum_{\langle i,j \rangle} D\, \boldsymbol{r}_{ij} \cdot (\boldsymbol{m}_i \times \boldsymbol{m}_j) - K \sum_j m_{i,z}^2$$

where $\langle i,j \rangle$ represent the nearest-neighbor bonds between $i,j$, $J$ is the exchange coupling, $D$ is the bulk DMI, $H$ and $K$ are field and anisotropy respectively. As mentioned above, the lattice is body-center-cubic to introduce an implicit anisotropy. The $z$-axis in the simulation is the (111) direction of bcc lattice, so the terms $J$ and $D$ act on nearest-neighbor on the bcc lattice (Fig. 11). Also, we did the simulation on thin film along the (111) direction. Since we want to mainly focus on the domain walls and helical configurations in our simulation, the $\phi$ shifting of $\overleftrightarrow{Q}$ is not considered here. Therefore, the six-order anisotropy term is not included.

The simulations we show are for $J = 1.0$, $D/J = 0.628$ and $KJ/D^2 = -0.14$. The value of K is chosen to match $q$, i.e., to move the orientation of the $\boldsymbol{Q}$'s toward (111). To understand the structure of domain walls (DWs) and their intersections, we chose initial configurations with specific textures, and relaxed the energy. Our phenomenological model and simulation method allow us to observe the stability, geometry and distortion of DWs and their intersections. To compare with SP-STM data, we look at the surface manifestation of textures whose energetics is governed by $\mathcal{H}$.

**Structural characterization of MnGe thin films**

The B20 structure of MnGe is comprised of alternating layers of Mn and Ge monomers and trimers with a specific stacking order and orientation as shown in Figure 12a-b. Along the (111) direction, the unit cell contains three repeats of the stacking sequence referred to as "quadruple layers" (QLs). *A priori*, there are sixteen unique surfaces possible, considering the four possible surface terminating layers (Mn-sparse, Ge-sparse, Mn-dense, and Ge-dense), two crystal chiralities, and two stacking sequences. The STM image in Fig. 12c-d was taken from the same area shown in Fig. 4 of the main text. Following the procedure for comparing experimental STM images and simulated images from density functional theory (DFT) we developed for B20 FeGe (*19*), we assign the bright spots in Fig. 12c to a Mn-sparse layer (Fig. 12d). Fainter contrast between these lattice points is attributed to the next underlying layer, assigned to a Ge-dense layer.

Figure 13 shows larger-scale characterization by atomic force microscopy (AFM), STM and x-ray diffraction (XRD). The film surface is primarily atomically smooth with terraces connected by atomic-scale steps and small voids (~ few nm deep).

**Spin dependent density of states**

To better understand the relationship between the stripe contrast and the spin-dependent density of states at the surface, spectroscopy was performed at points where the tip's magnetization was aligned with the sample's or anti-aligned (dark). These points are labeled with spots in the difference STM image. Both spectra overlap in the region of low bias (+/- 20mV) and begin to deviate at higher positive and negative bias. This difference in intensity is due to the contribution of the local spin polarization to the tunneling conductance, and underlies the spin-dependent contrast in the spatial dI/dV images.

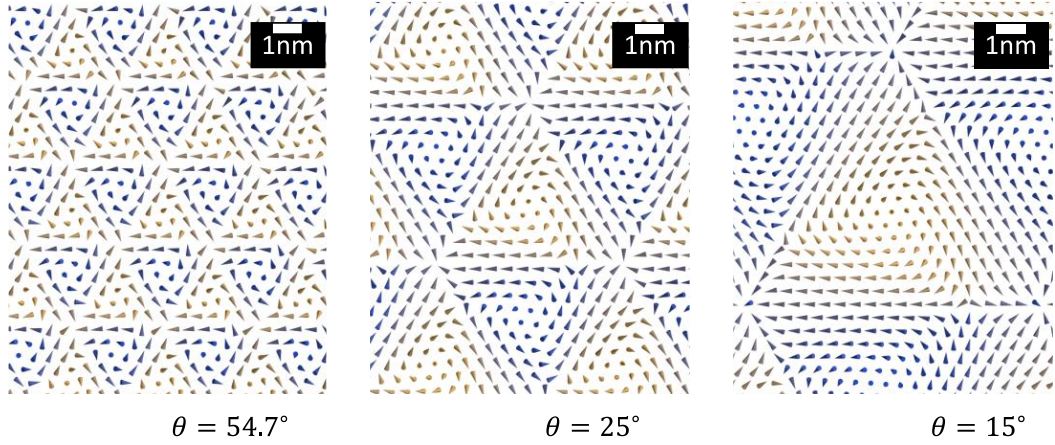

$\theta = 54.7°$      $\theta = 25°$      $\theta = 15°$

**Fig. S1.**

Spin textures on the (111) surface of a sample whose bulk is a 3Q hedgehog crystal. Here $\theta$ is the angle between each of the three $\widehat{Q}_i$ vectors and (111) the surface normal. The left panel $\theta = 54.7°$ corresponds to the three $\widehat{Q}_i$'s along (100), while decreasing $\theta$ brings them closer to (111).

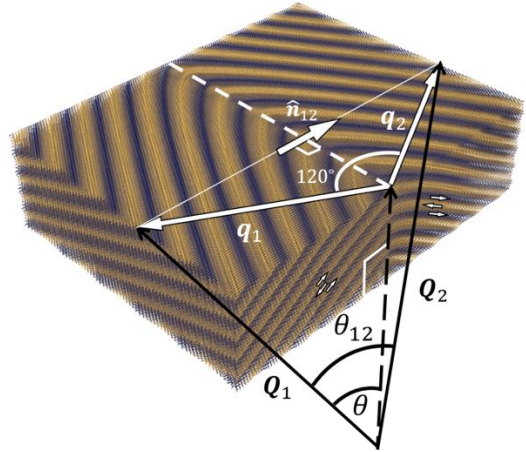

**Fig. S2.**

Domain wall between helical domains with wave-vectors $\mathbf{Q}_1$ and $\mathbf{Q}_2$. Note the relation between the 3D wavevector $\mathbf{Q}_i$ and its 2D projection $\mathbf{q}_i$. Also shown is the angle $\theta$ between each $\widehat{\mathbf{Q}}_i$ and the surface normal $\hat{\mathbf{z}} = (111)$, the angle $\theta_{12}$ between $\mathbf{Q}_1$ and $\mathbf{Q}_2$, and the normal vector $\hat{\mathbf{n}}_{12}$ to domain wall.

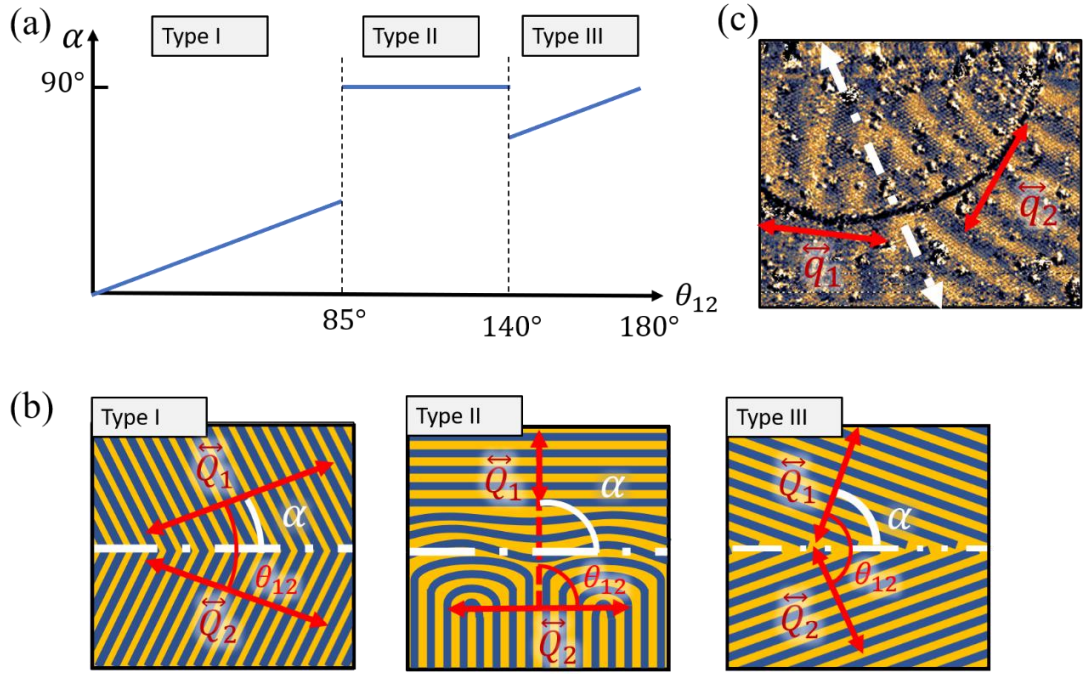

**Fig. S3.**
(a) Summary of the results of ref. (*28*), showing the region of stability of different types of DWs and the relation between $\theta_{12}$ (angle between $Q_1$ and $Q_2$) and $\alpha$ (angle between $Q_1$ and the DW).
(b) Illustrations of the three types of DWs showing $Q_1$, $Q_2$, helical stripes, and the angles $\theta_{12}$ and $\alpha$. Both $Q_i$'s lie in the plane of the figure. (c) SP-STM data shows a type-I DW with an angle $\angle q_1 q_2 \simeq 120°$ between the in-plane projection of helical wave vectors.

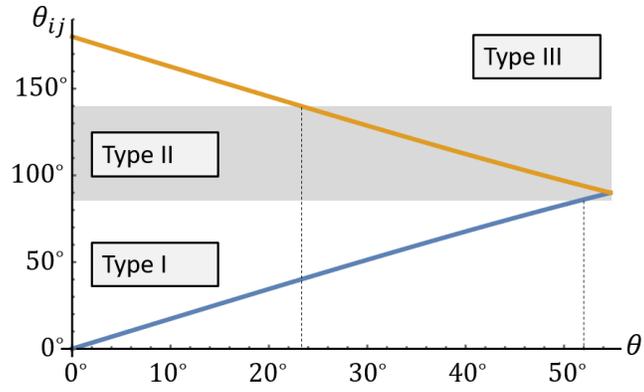

**Fig. S4.**
Solutions of eq. (2) for the angle $\theta_{ij}$ as a function of $\theta$ (see text and Fig. 2 for definitions of these angles).

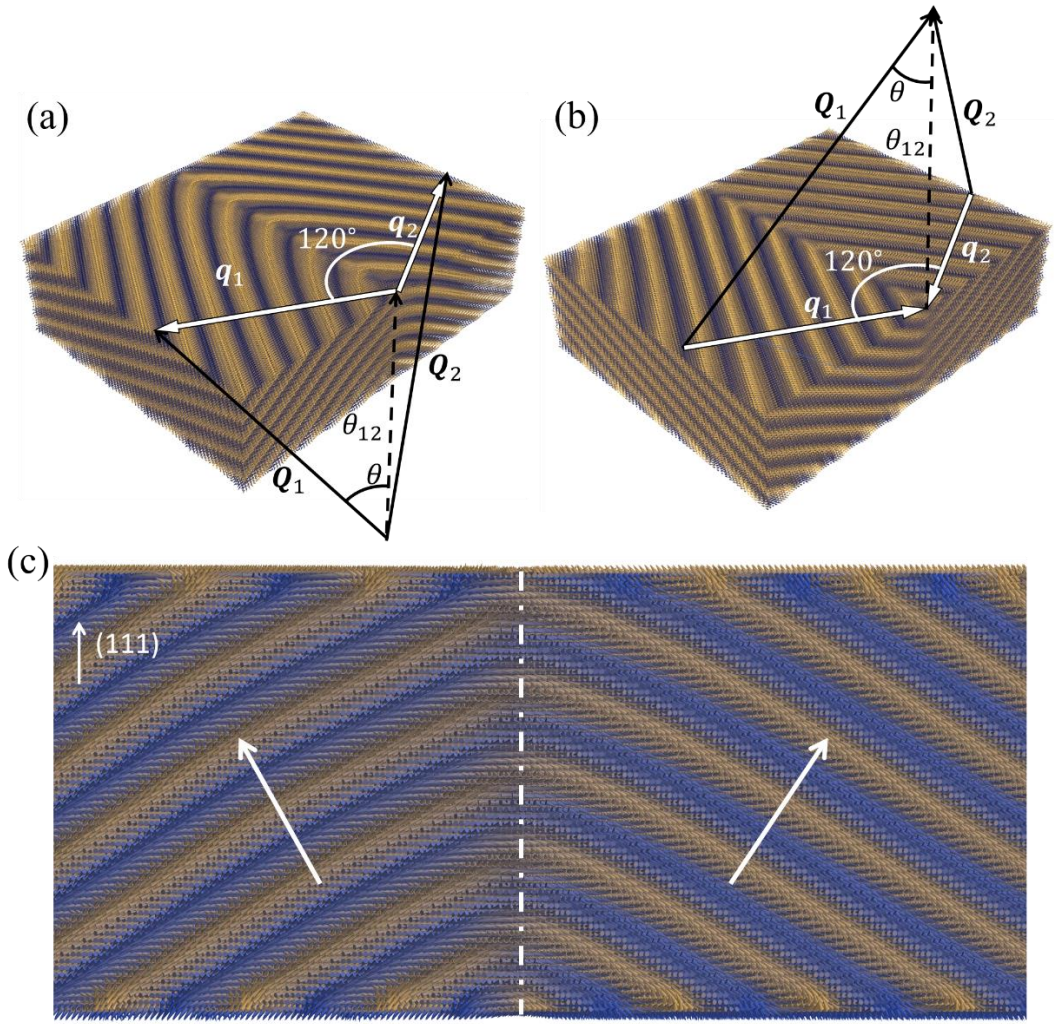

**Fig. S5.**

(a) When the 3D wave vectors $Q_1$ and $Q_2$ point away from the DW, one observes rounding of the helical stripes at the DW on the surface. (b) When $Q_1$ and $Q_2$ point toward the DW, one observes a sharpening at the DW on the surface. (c) Cross-sectional view for the helices and DW. The distortion near the surface reveals the surface effect.

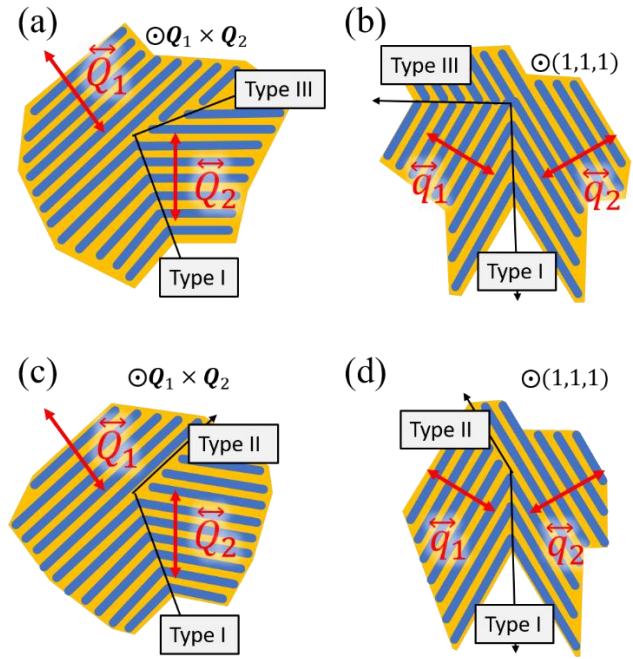

**Fig. S6.**

Intersection of two domain walls. (a) shows the position of type I and type III DWs from a view along $Q_1 \times Q_2$. (b) shows the position of type I and type III DWs from the top view along (111). (c) shows the position of type I and type III DWs from a 3D view.along $Q_1 \times Q_2$. (d) shows the position of type I and type III DWs from a view.along (111). Here we assume the case that type II and type III DWs exist in the system. The illustration (c) and (d) shows the special case of type II DWs for convenience, but the relation between $Q$'s domains and DWs are the same for the zig-zag kind of type II DWs.

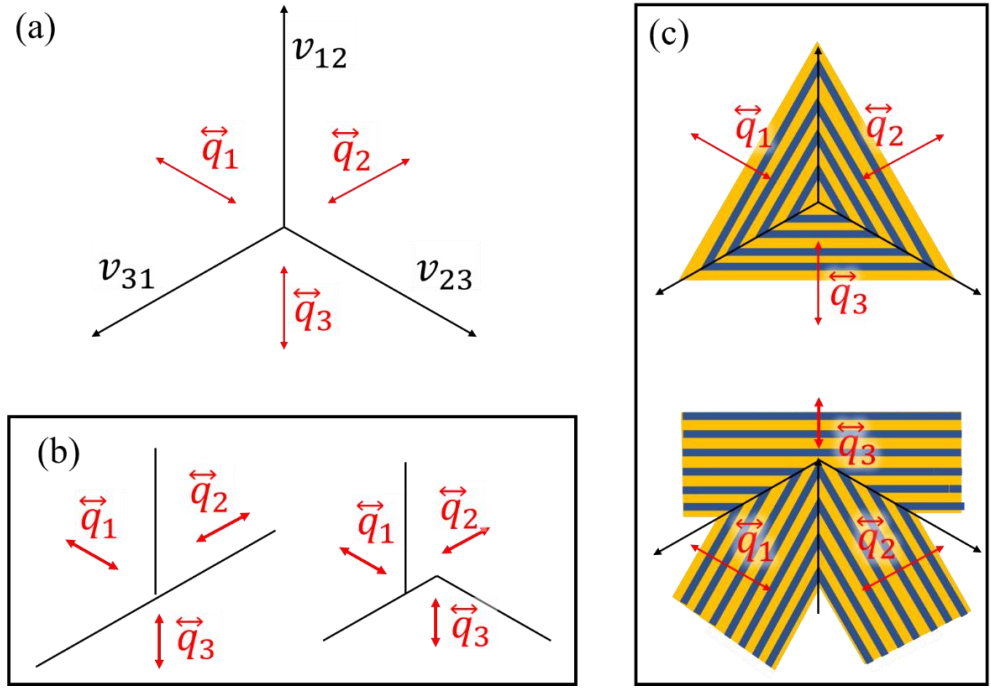

**Fig. S7.**

(a) shows the allowed directions of DWs for domains $Q_1$, $Q_2$ and $Q_3$. (b) shows two examples of intersection which are not allowed since the type I DW should be bisectors even for the view from top. (c) shows the only two allowed types of intersection in our system for type I DWs. They are related to different permutations of domains about the intersections.

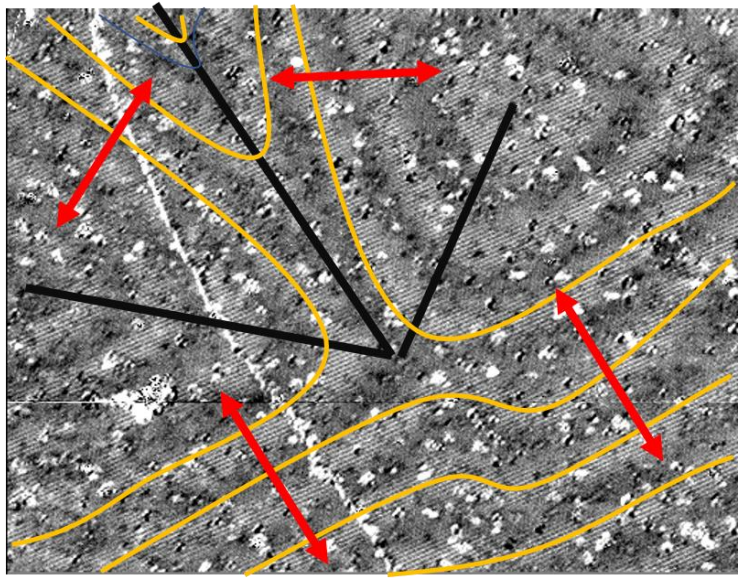

**Fig. S8.**
π-texture intersection seen by SP-STM on MnGe.

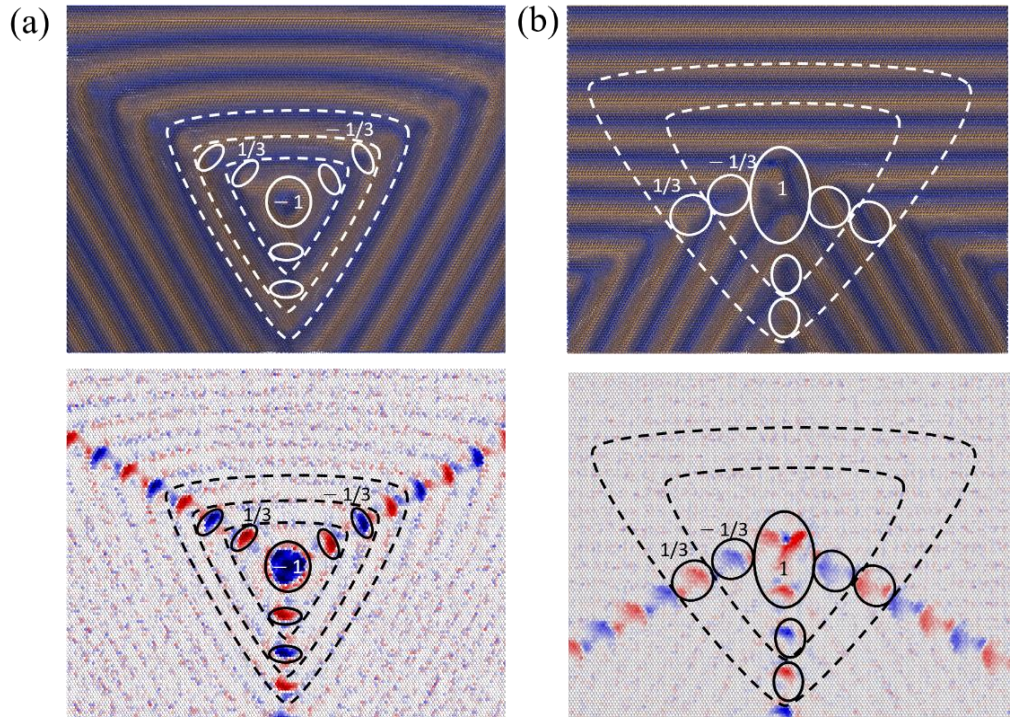

**Fig. S9.**

(a) Target texture and (b) $\pi$ texture (upper panels) with their topological charge density (lower panels) which is concentrated at the DWs. Except at the core, there is an alternating topological charge with magnitude 1/3 at the intersection of the DW with the stripes.

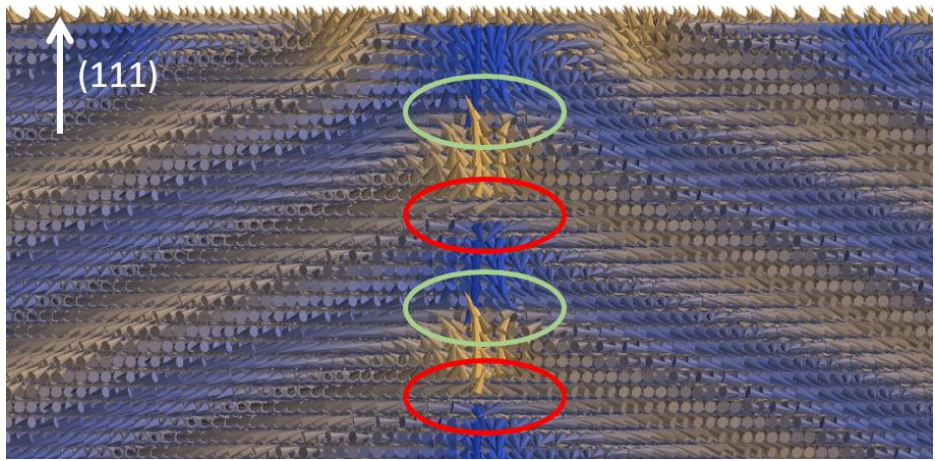

**Fig. S10.**

Cross-sectional view of target texture showing a chain of hedgehogs and anti-hedgehogs along the $\hat{z} = (111)$ axis in the core of the texture.

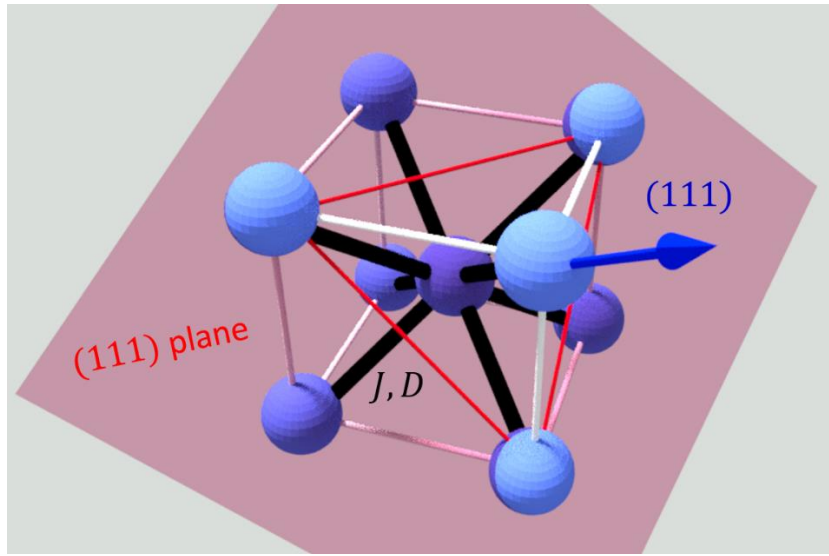

**Fig. S11.**

The bcc lattice used for our micromagnetic simulations. The interactions act on the nearest-neighbor bonds (black bonds). The (111) direction of the bcc lattice is set as the $z$-direction, so spins on the (111) plane form triangular lattice for each layer in the simulation.

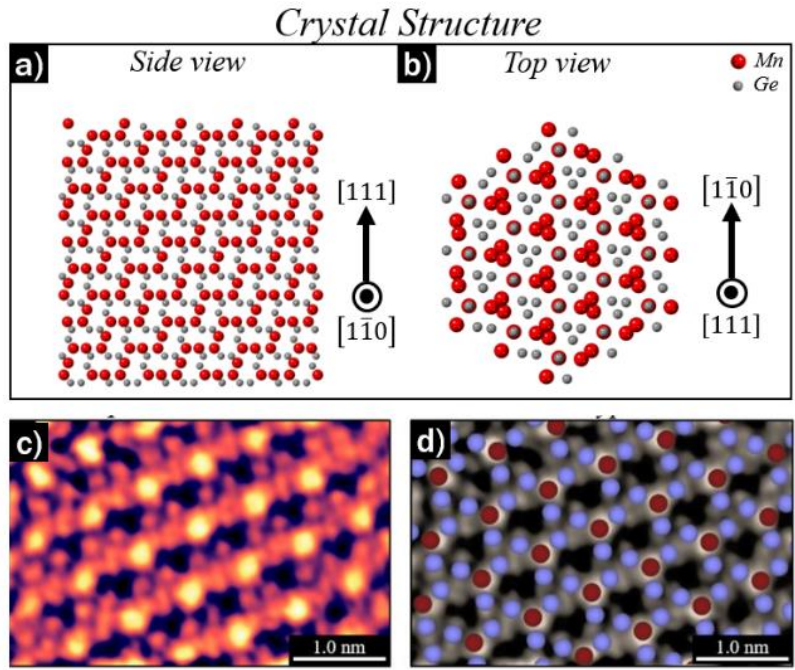

**Fig. S12.**
Atomic structure of B20 MnGe. (a-b) crystal models showing the layered B20 structure with alternating sparse and dense layers of Mn and Ge atoms (c) atomic resolution STM image taken in the region of Fig. 4 of the main text. (d) Overlay of atomic lattice assignment. Red spots are Mn atoms of the Mn-s surface, purple spots are the next layer of Ge-d.

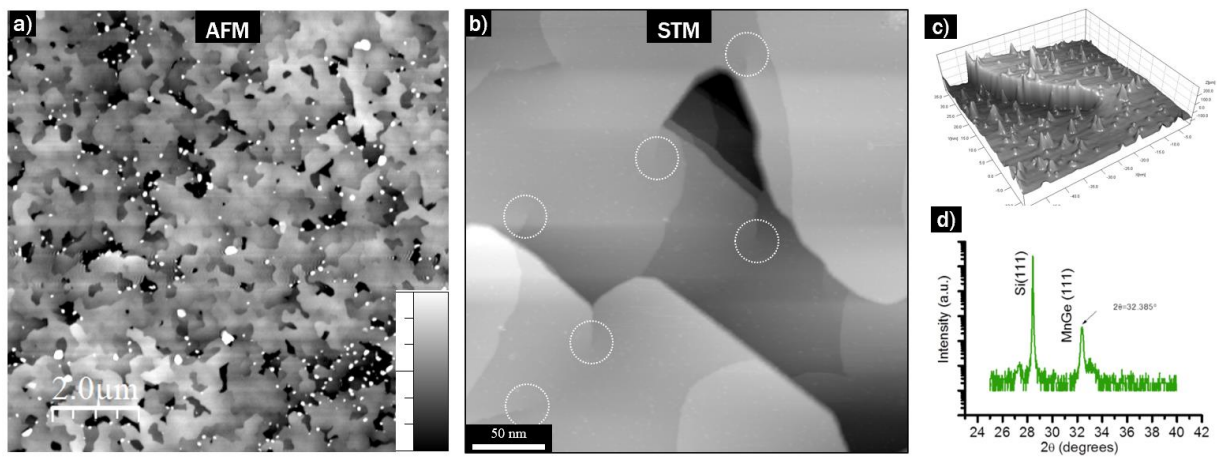

**Fig. S13.**
(a) AFM image of 200 nm MnGe film. The gray scale represents 10nm in height difference. (b) Large scale STM image of 80 nm MnGe film. Dotted circles indicate screw dislocations in the film. (c) higher magnification image showing the surface wrapping about a screw dislocation. (d) X-ray diffraction of the 80 nm MnGe film. The (111) reflection gives 2.7615 Å, compared to the expected bulk value of 2.768 Å (~ 0.2% reduction).

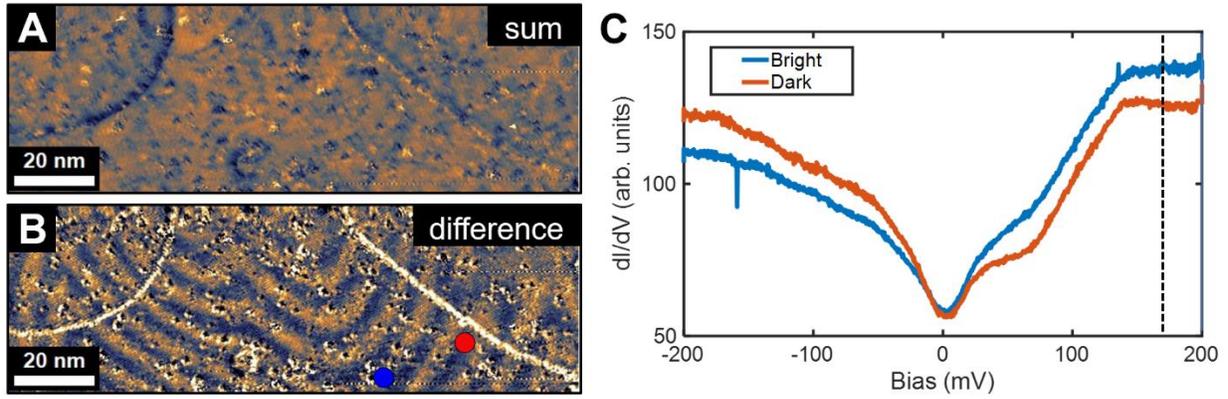

**Fig. 14.**

Sum/difference SP-STM images corresponding to the raw data in Fig. 2 of the main text. (a) sum of images with opposite tip polarizations, confirming no magnetic contrast. (b) difference image as in Fig. 2. (c) dI/dV spectra with the tip positioned at bright / dark regions for the tip polarization as in Fig. 2c.